\documentclass[letterpaper,titlepage,11pt]{article}

\pdfoutput=1

\usepackage{amssymb,amsmath,amsfonts}
\usepackage{epsfig}
\usepackage{ccaption}
\usepackage{graphicx}

\usepackage[
      colorlinks=true,
      linkcolor=blue,
      urlcolor=blue,
      filecolor=black,
      citecolor=red,
      pdfstartview=FitV,
      pdftitle={},
        pdfauthor={Roberto Emparan, Ryotaku Suzuki, Kentaro Tanabe},
        pdfsubject={},
        pdfkeywords={},
        pdfpagemode=None,
        bookmarksopen=true
      ]{hyperref}

\def\clock{{\count0=\time
           \divide\count0 60
           \ifnum\count0<10 0\fi\the\count0
           \multiply\count0 -60 \advance\count0 \time
           :\ifnum\count0<10 0\fi \the\count0
         }}
\newcommand{\timestamp}{{\small\vbox{\hbox{\tt\jobname.tex}
\hbox{\the\day/\the\month/\the\year, \clock}}}}

\newcommand{\ie}{{\it i.e.,\,}}
\newcommand{\eg}{{\it e.g.,\,}}

\newcommand{\lp}{\left(}
\newcommand{\rp}{\right)}

\newcommand{\mc}[1]{\mathcal{#1}}

\newcommand{\beq}{\begin{equation}}
\newcommand{\eeq}{\end{equation}}
\newcommand{\bea}{\begin{eqnarray}}
\newcommand{\eea}{\end{eqnarray}}
\newcommand{\beqa}{\begin{eqnarray}}
\newcommand{\eeqa}{\end{eqnarray}}

\newcommand{\sR}{\mathsf{R}}

\numberwithin{equation}{section}

\begin{document}

\begin{titlepage}
\leftline{}
\vskip 1cm
\centerline{\LARGE \bf Instability of rotating black holes:}\medskip
\centerline{\LARGE \bf large $D$ analysis}
\vskip 1.6cm
\centerline{\bf Roberto Emparan$^{a,b}$, Ryotaku Suzuki$^{c}$, Kentaro Tanabe$^{b}$}
\vskip 0.5cm
\centerline{\sl $^{a}$Instituci\'o Catalana de Recerca i Estudis
Avan\c cats (ICREA)}
\centerline{\sl Passeig Llu\'{\i}s Companys 23, E-08010 Barcelona, Spain}
\smallskip
\centerline{\sl $^{b}$Departament de F{\'\i}sica Fonamental, Institut de
Ci\`encies del Cosmos,}
\centerline{\sl  Universitat de
Barcelona, Mart\'{\i} i Franqu\`es 1, E-08028 Barcelona, Spain}
\smallskip
\centerline{\sl $^{c}$Department of Physics, Osaka City University, Osaka 558-8585, Japan}
\vskip 0.5cm
\centerline{\small\tt emparan@ub.edu,\, ryotaku@sci.osaka-cu.ac.jp,\, ktanabe@ffn.ub.es}

\vskip 1.6cm
\centerline{\bf Abstract} \vskip 0.2cm \noindent 
We study the stability of odd-dimensional rotating black holes with equal angular momenta by performing an expansion in the inverse of the number of dimensions $D$. Universality at large $D$ allows us to calculate analytically the complex frequency of quasinormal modes to next-to-leading order in the expansion. We identify the onset of non-axisymmetric, bar-mode instabilities at a specific finite rotation, and axisymmetric instabilities at larger rotation. The former occur at the threshold where the modes become superradiant, and before the ultraspinning regime is reached. Our results fully confirm the picture found in numerical studies, with very good quantitative agreement. We extend the analysis to the same class of black holes in Anti-deSitter space, and find the same qualitative features. We also discuss the appearance at high frequencies of the universal set of (stable) quasinormal modes.

\end{titlepage}
\pagestyle{empty}
\small
\normalsize
\newpage
\pagestyle{plain}
\setcounter{page}{1}

\section{Introduction and summary of results}

The equilibrium and stability of most gravitating systems is dominated by the antagonistic pull of gravitational and centrifugal forces. This balance plays out most starkly in rotating black holes, which makes the study of their stability an important part of the dynamics of General Relativity.
A useful strategy for gaining insight into this problem is to let the number of dimensions $D$ vary. By tuning $D$ as a parameter in the theory we can alter the relative balance between the forces at play: increasing $D$ reduces the range of the gravitational interaction, but leaves unchanged the inverse-square fall-off of the centrifugal potential~\cite{Emparan:2008eg}. 

In $D=4$ the stability of Kerr black holes\footnote{Throughout this paper we only consider linear mode stability.} against centrifugal disruption appears to be secured by the fact that they stop to exist as equilibrium systems beyond a certain angular momentum. 
This safeguard, however, disappears for the Myers-Perry (MP) black holes in $D\geq 6$ \cite{Myers:1986un}, which admit regimes of arbitrarily large angular momenta, so ref.~\cite{Emparan:2003sy} argued (invoking also other arguments) that these solutions must become unstable for large enough rotation. Later studies have confirmed the existence of these instabilities, in the form of axisymmetric perturbations of these black holes \cite{Dias:2009iu,Dias:2010eu,Dias:2010maa,Dias:2010gk,Dias:2011jg,Cardoso:2013pza}, and also of non-axisymmetric (`bar-mode') instabilities that set in at lower values of the rotation and therefore are presumably the dominant mechanism in the destabilization of fastly-rotating black holes \cite{Shibata:2009ad,Shibata:2010wz,Hartnett:2013fba}.\footnote{See also \cite{Armas:2010hz}.} All these works make heavy use of numerical techniques for solving Einstein's equations for the perturbed black holes. Our purpose here is to begin the investigation of the problem by analytic means using the large $D$ expansion as developed in \cite{Asnin:2007rw,Emparan:2013moa,Emparan:2013xia,Emparan:2014cia}.


The problem is rendered simpler for black holes in odd spacetime dimension 
\beq
D=2N+3\,,\qquad N=1,2,\dots,
\eeq
when the $N+1$ independent angular momenta are all non-zero and equal. In this case, even though the geometry is not spherically symmetric, it depends non-trivially only on the radial coordinate (\ie\ has cohomogeneity 1). Unlike the case with a single spin, the angular momentum of these black holes is bounded above, but a thermodynamic argument suggests that dynamic instabilities may be found within the range \cite{Dias:2010eu}
\beq\label{instwin}
\frac1{\sqrt{2}}<\frac{a}{r_+}\leq \sqrt{\frac{N}{N+1}}\,.
\eeq
Here $a$ is the rotation parameter and $r_+$ the horizon radius, and the upper bound is the extremal limit. The lower bound marks the onset of the `ultra-spinning regime' defined as the threshold for the appearance of negative eigenvalues of the Hessian  $-\partial^2 S/(\partial J_i\partial J_j)$. At this rotation an axisymmetric zero-mode perturbation exists that deforms the black hole along the family of MP solutions, so this is not associated to the onset of a dynamic instability; however, unstable behavior is expected to set in at some larger value of the rotation. 
Since the thermodynamic Hessian is sensitive only to axisymmetric perturbations, it is of interest to determine whether non-axisymmetric instabilities comply with this bound on the rotation or not. For singly-spinning black holes this question was answered in the negative by the numerical analyses of 
refs.~\cite{Shibata:2009ad,Shibata:2010wz}.

The putative instability window \eqref{instwin} is open only for $N>1$, in accord with the stability results of \cite{Murata:2008yx} for these black holes in five dimensions. For $N=2,3$, a numerical study in ref.~\cite{Dias:2010eu} found the expected axisymmetric instabilities, all lying within the range \eqref{instwin}. Perhaps less expected is the recent numerical find of non-axisymmetric instabilities for $N=2,3,4,5,6$ (and not for $N=1$) at rotations \textit{lower} than the ultraspinning range \eqref{instwin} \cite{Hartnett:2013fba}. Moreover, these unstable modes are superradiant, with frequencies satisfying
\beq\label{suprad}
\mathrm{Re}\,\omega< m \Omega_H\,,
\eeq
($\Omega_H$ is the horizon angular velocity) so the black hole can radiate them away while increasing its horizon area.

We have succeeded in reproducing and extending these numerical findings through an analytical calculation of quasinormal frequencies up to next-to-leading order in an expansion in $1/N$.


\paragraph{MP black holes.} We have found linearized gravitational scalar perturbations that become unstable whenever the rotation is larger than the critical value $a=a_c$ given by
\beq\label{crita}
\frac{a_c}{r_+}=\sqrt{1-\frac1{\ell}}\lp 1-\frac{1}{N} \frac{m^2}{4\ell^2}\rp+\mathcal{O}\lp N^{-2}\rp\,,
\eeq
where $\ell\geq 2$ is the angular momentum number, for any non-zero magnetic number $m$, with $\ell-|m|=2\kappa$ 
an even number.\footnote{Other references, \eg\ \cite{Dias:2010eu,Durkee:2010ea,Hartnett:2013fba,Cardoso:2013pza}, employ $\kappa$ instead of $\ell$.} The dominant mode, \ie\ with the smallest $a_c$, has $\ell=m=2$ and therefore we find
\beq\label{naxibound}
\frac{a}{r_+}>\frac1{\sqrt{2}}\lp 1-\frac{1}{4N}\rp+\mathcal{O}\lp N^{-2}\rp\quad\Rightarrow\quad \mathrm{non\mbox{-}axisymmetric~instability}\,.
\eeq
Observe that this non-axisymmetric, bar-mode instability sets in at a value of the rotation \textit{lower} than the ultraspinning bound in \eqref{instwin}. 

The instability of the black hole corresponds to a change from negative to positive value of the imaginary part of a complex quasinormal frequency $\omega$. At this critical rotation where $\mathrm{Im}\,\omega=0$, we find that the real part of the mode frequency becomes
\beq
\omega=m\frac{a_c}{r_+^2}
\eeq
which sits precisely at the threshold for superradiance \eqref{suprad}.

We also find axisymmetric modes, with $m=0$, that become unstable with a critical rotation \eqref{crita}. However, the mode $(\ell,m)=(2,0)$ corresponds merely to a perturbation that adds angular momentum along the MP family of solutions --- indeed we correctly find that its critical rotation parameter is the same as the threshold for the ultraspinning regime \eqref{instwin}. The first unstable axisymmetric mode has $\ell=4$ and thus we get
\beq\label{axibound}
\frac{a}{r_+}>\frac{\sqrt{3}}2+\mathcal{O}\lp N^{-2}\rp\quad\Rightarrow\quad \mathrm{axisymmetric~instability}\,.
\eeq

These axisymmetric modes are purely imaginary, so at the critical rotation $a_c$ they become stationary zero modes. These are expected to mark the appearance of new branches of solutions in the generic manner indicated in \cite{Emparan:2003sy,Emparan:2007wm}.

\begin{figure}[t]
 \begin{center}
  \includegraphics[width=.47\textwidth,angle=0]{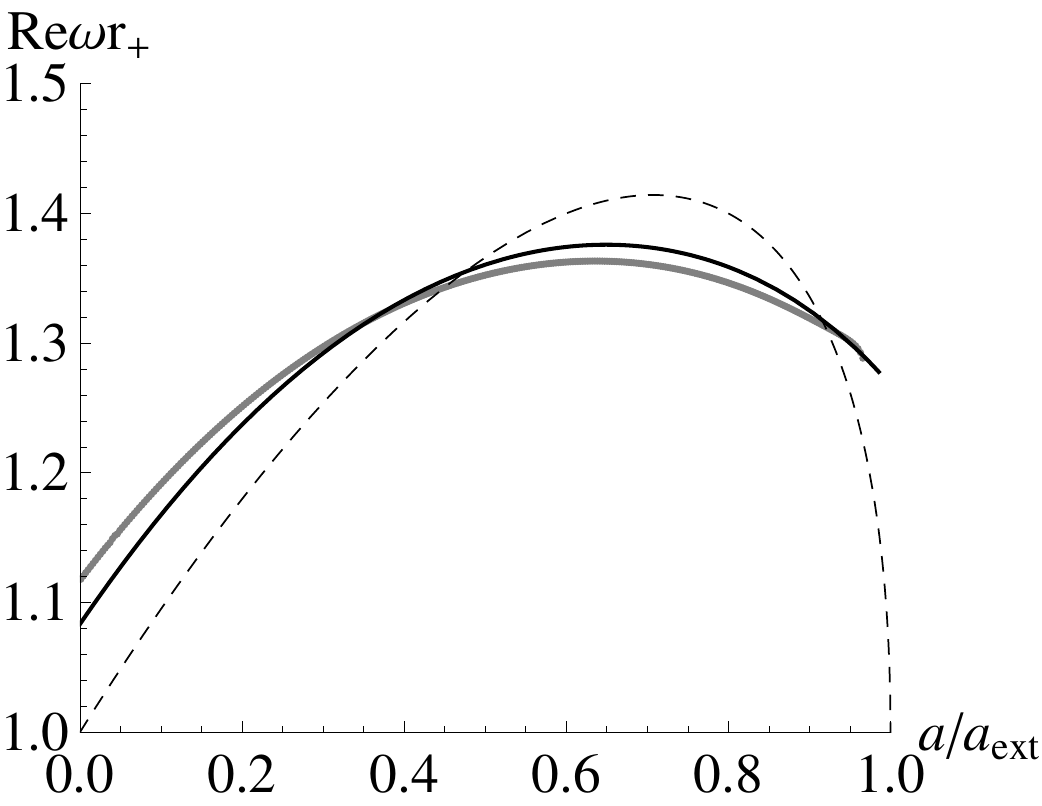}
  \hspace{5mm}
  \includegraphics[width=.47\textwidth,angle=0]{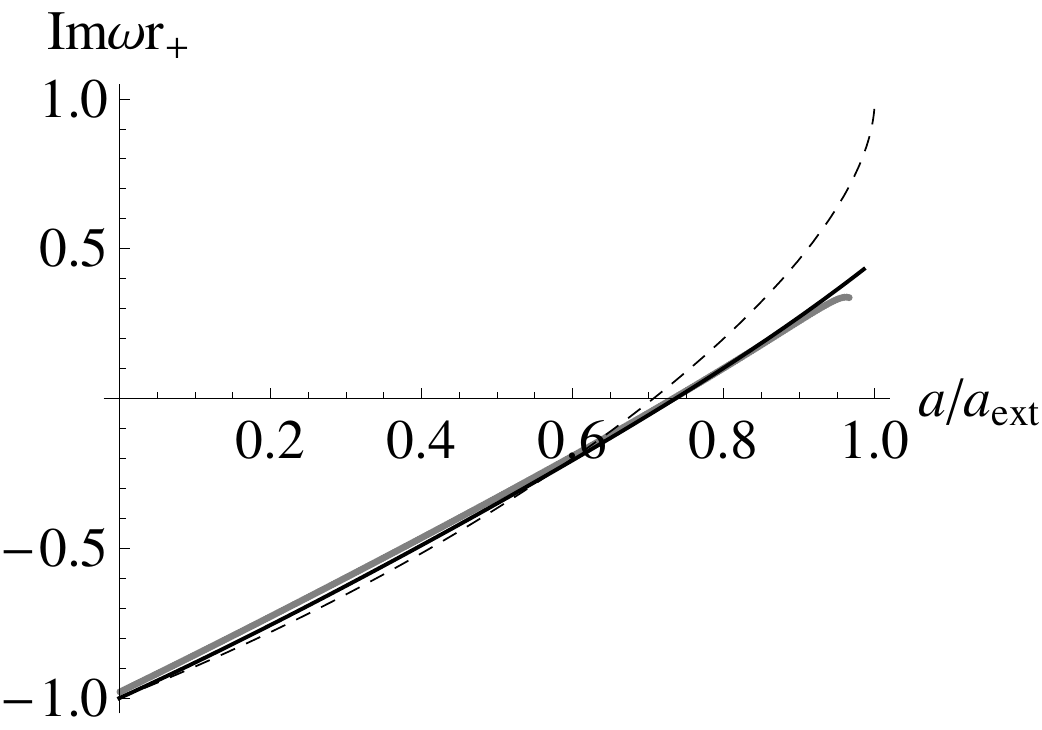}
   \end{center}
 \vspace{-5mm}
 \caption{\small Comparison between analytical and numerical calculations of the 
real and imaginary frequencies of the dominant unstable quasinormal mode with $\ell=m=2$ for $N=6$ ($D=15$). Black lines: analytical result to next-to-leading order in $1/N$. Gray lines: numerical results of \cite{Hartnett:2013fba} for $N=6$. Dashed lines: analytical result to leading order.}
 \label{fig:comp}
\end{figure}

Quantitative comparison with the numerical results of \cite{Hartnett:2013fba} for the critical values $a_c/r_+$ shows very good agreement, with typical differences of size smaller than $\approx 1/{(2N)}^2$, although the accuracy worsens for values of $a_c$ closer to extremal rotation, and in particular for axisymmetric modes. More generally, the frequency spectrum $\omega_{\ell,m}$ as a function of $a$ is well reproduced, as illustrated in fig.~\ref{fig:comp} for the dominant unstable mode $\ell=m=2$ in $D=15$.
%

\paragraph{Rotating AdS black holes.} We have extended the leading-order results to rotating black holes in Anti-deSitter space, for which the extremality bound at $N\to\infty$ is\footnote{Our definition of radial coordinate, and hence of $r_+$, is the same as in \cite{Kunduri:2006qa}. To relate it to the one in \cite{Gibbons:2004js,Gibbons:2004ai} see app.~\ref{app:hessian}.}
\beq\label{adsext}
\frac{a_\mathrm{ext}}{r_+}=\frac1{\sqrt{1+\frac{r_+^2}{L^2}}}+\mathcal{O}\lp N^{-1}\rp
\eeq
(note that $a_\mathrm{ext}<L$). In appendix~\ref{app:hessian} we compute the ultraspinning bound that results from the thermodynamic negative modes (exactly in $N$), yielding
\beq\label{ultraads}
\frac{a_\mathrm{ultra}}{r_+}=\frac1{\sqrt{2+\frac{r_+^2}{L^2}}}\,.
\eeq

Our conclusions regarding dynamical stability are qualitatively the same as above, with \eqref{crita} replaced by
\beq\label{critaads}
\frac{a_c}{r_+}
=\sqrt{\frac{\ell-1+\frac{r_+^2}{L^2}}{\lp \ell+\frac{r_+^2}{L^2}\rp\lp 1+\frac{r_+^2}{L^2}\rp}}+\mathcal{O}\lp N^{-1}\rp
\,,
\eeq
and the superradiance threshold being at
\beq
\omega=a m \lp 1+\frac{r_+^2}{L^2}\rp+\mathcal{O}\lp N^{-1}\rp\,.
\eeq
Again the dominant instability occurs for $\ell=m=2$, for which $a_c=a_\mathrm{ultra}$, so these black holes are unstable whenever
\beq\label{domaads}
a_\mathrm{ultra}+\mathcal{O}\lp N^{-1}\rp<a\leq a_\mathrm{ext}\,.
\eeq
To this order we find that the superradiant, and stability bounds  coincide with the ultraspinning bound, but we expect that $1/N$ corrections drive the onset of the instability below the ultraspinning bound. Axisymmetric instability sets in at the critical rotation \eqref{critaads} with $\ell=4$.

Observe that $a_c/a_\mathrm{ext}$ grows with $r_+/L$: the negative cosmological constant increases the range of stability of the black holes, as might have been expected. Nevertheless, the instability does not disappear. In particular, if we fix the black hole mass and increase its angular momentum, then we always encounter an instability no matter how large the mass is.\footnote{Closeness to extremality may make the conclusions in this regime less reliably applicable for finite $N$.}

\paragraph{Schwarzschild vs.\ MP~(-AdS) quasinormal modes.}
Ref.~\cite{Emparan:2013xia} found that in the leading large $D$ limit the near-horizon geometry of MP black holes is locally a boost of the near-horizon Schwarzschild solution. In the present case this boost has velocity  $a/r_+$ homogeneously over the horizon. Naively this implies that  by boosting the leading large $D$ quasinormal frequencies of Schwarzschild we obtain quasinormal MP frequencies of the form
\beq\label{schboost}
\omega= m\frac{a}{r_+^2} +\omega_\text{Sch}\sqrt{1-\frac{a^2}{r_+^2}}\,.
\eeq
If the $a$-dependence of all MP quasinormal frequencies were of this form, then no instability could appear at finite rotation (to leading large $D$ order, at least). None of those modes would be superradiant, either. A similar relationship exists with black holes in AdS, so the mode frequencies with rotation and/or in AdS would follow from the ones in Schwarzschild (without a cosmological constant) as
\beq
\omega= m\frac{a}{r_+^2}\lp 1+\frac{r_+^2}{L^2}\rp  +\omega_\text{Sch}\sqrt{1-\frac{a^2}{r_+^2}-\frac{a^2}{L^2}}\,.
\eeq

However, these equations are not valid for all the quasinormal spectrum. They apply only if the condition that determines the quasinormal frequencies knows only about the leading order near-horizon geometry. This is in fact true for several classes of modes, including all those with $\omega r_+=\mathcal{O}(D)$ --- so the `universal quasinormal modes' in \cite{Emparan:2014cia} appear in boosted form for MP black holes --- and also for some of the modes with frequencies $\omega r_+=\mathcal{O}(D^0)$. 
However, we have found a class of modes determined by equations and boundary conditions at the horizon that involve structure beyond a boosted version of the leading order Schwarzschild solution. The $a$-dependence of $\omega$ is then more complicated than \eqref{schboost}. All unstable modes fall in this category.

\bigskip

The remainder of the article is organized as follows: the next section describes the large $D$ limit of the solution and the crucial boost relation between leading order near-horizon geometries. Secs.~\ref{sec:perts} and \ref{sec:qnman} are mostly technical sections describing the calculation of quasinormal modes in the large $N$ expansion. The impatient reader may jump to the main result of this analysis, eq.~\eqref{QNMeq}, which yields the quasinormal frequency spectrum to leading order in the expansion. In sec.~\ref{sec:instab} we study the instabilities that result from this spectrum. In sec.~\ref{sec:adsmp} we extend the study to AdS rotating black holes. Sec.~\ref{sec:odmodes} shows how modes with high frequencies $\omega r_+=\mc{O}(D)$ are simply boosted versions of the universal quasinormal modes of Schwarzschild. We make some final remarks in sec.~\ref{sec:fin}. We have put in appendices a number of technical steps for the main calculations, but other appendices may be of more general interest: app.~\ref{app:hessian} provides an analysis of thermodynamic negative modes of rotating AdS black holes; app.~\ref{app:sncpn} describes the relation between harmonics of $S^{2N+1}$ and harmonics of $\mathbb{CP}^N$; app.~\ref{app:ell0} proves the stability of modes $\ell=0$ at any $N$. A \textit{Mathematica} file attached to the article contains results of the next-to-leading order calculation of quasinormal frequencies that are too lengthy to include here.

\bigskip 

\noindent\textbf{Note:} Dias, Hartnett, and Santos have kindly informed us of their recent numerical calculation of quasinormal frequencies of asymptotically flat rotating black holes, including large values of $D$, with good agreement with our results~\cite{DHS}.

\section{The metric and its large $D$ limit}
\label{sec:metrD}

The metric of the $D=2N+3$ dimensional MP black hole with equal angular momenta can be written in the form \cite{Kunduri:2006qa}
%
\begin{eqnarray}\label{MPbh}
ds^{2} &=& -\frac{G(r)}{H(r)}dt^{2} +\frac{dr^{2}}{G(r)} 
 +r^2 H(r) \lp d\psi -\Omega(r) dt+A_{a}dx^{a}\rp^{2} +r^{2}\hat{g}_{ab}dx^{a}dx^{b},
\end{eqnarray}
%
where
%
\begin{eqnarray}
&&
G(r) =  1-\left(\frac{r_{0}}{r}\right)^{2N}\lp 1-\frac{a^2}{r^2}\rp \,,\\
&&
H(r) = 1+\frac{a^{2}}{r^2}\left(\frac{r_{0}}{r}\right)^{2N}, \\
&&
\Omega(r) = \frac{a}{r^2 H(r)} \left(\frac{r_{0}}{r}\right)^{2N}\,,
\end{eqnarray}
and $\hat{g}_{ab}$ is the Fubini-Study metric on $\mathbb{CP}^{N}$ with Kahler potential $A_{a}$. We take non-negative rotation $a\geq 0$ without loss of generality. The event horizon is at the largest positive root $r=r_+$ of $G(r)$,\footnote{Note that $r_+$ is invariantly defined as the size of the $\mathbb{CP}^N$ factor of the horizon.}. Its angular velocity is
\beq
\Omega_H=\frac{a}{r_+^2}\,,
\eeq
and its surface gravity
\beq\label{surfg}
\kappa=\left. \frac{G'}{2\sqrt{H}}\right|_{r=r_+}=N \frac{r_0^N}{r_+^{N+1}}\lp 1 -\frac{N+1}{N}\frac{a^2}{r_+^2}\rp\,.
\eeq
In the extremal limit $a=a_\text{ext}$ the surface gravity is zero.

When $N$ is large we have
\beq
r_+\simeq r_0\lp 1-\frac{a^2}{r_0^2}\rp^{1/(2N)}\,,
\eeq
so, since we will remain away from the extremal limit, $a<(1-\mc{O}(N^{-1}))a_\mathrm{ext}=r_0$ (when $N\to\infty$), we get 
\beq
r_+= r_0\lp 1+\mc{O}\lp N^{-1}\rp\rp\,.
\eeq
Hereafter we set
\beq
r_0=1\,.
\eeq

As explained in \cite{Emparan:2013moa}, in the large $D$ limit at any finite distance outside the horizon, $r>1$, the far-zone metric is that of flat space. For the near-zone
we introduce the coordinate\footnote{To leading order in $1/N$, $\sR\simeq (r/r_+)^{2N}\simeq r^{2N}/(1-a^2)$. We could define $\sR$ by any of the latter expressions without altering the remainder of this section, but it would change details (though not conclusions) of the perturbation analysis at higher order.}
%
\begin{eqnarray}
\sR=r^{2N}\lp 1-\frac{a^2}{r^2}\rp^{-1}
\end{eqnarray}
%
in terms of which, to leading order at large $N$, the metric becomes
%
\begin{eqnarray}\label{nearmp}
ds^{2} &=& \frac{1}{4N^{2}}\frac{d\sR^{2}}{\sR(\sR-1)}-\left(1-\frac{\cosh^{2}{\alpha}}{\sR}\right)dt^{2}
+\left( 1+\frac{\sinh^{2}{\alpha}}{\sR} \right)\left( d\psi +A_{a}dx^{a}  \right)^{2} \notag \\ &&-\frac{2\sinh{\alpha}\cosh{\alpha}}{\sR}dt(d\psi+A_{a}dx^{x}) 
+\hat{g}_{ab}dx^{a}dx^{b},
\end{eqnarray}
%
where
%
\begin{eqnarray}
\tanh{\alpha} = a\,.
\end{eqnarray}
%
This metric can be obtained from the near-zone metric of the Schwarzschild black hole ($\alpha=0$)
%
\beq\label{nearsch}
ds^{2} = \frac{1}{4N^{2}}\frac{d\sR^{2}}{\sR(\sR-1)}-\left(1-\frac{1}{\sR}\right)dt^{2} +\left( d\psi +A_{a}dx^{a}  \right)^{2} 
+\hat{g}_{ab}dx^{a}dx^{b}\,,
\eeq
%
by performing the frame transformation 
%
\beqa
dt\rightarrow dt\cosh{\alpha} -(d\psi+A_{a}dx^{a})\sinh{\alpha},\label{transL1} \\
d\psi +A_{a}dx^{a}  \rightarrow (d\psi+A_{a}dx^{a})\cosh{\alpha} -dt\sinh{\alpha},  
\label{transL2}
\eeqa
%
or, as a vector basis transformation,
%
\begin{gather}
\partial_{t} \rightarrow \cosh{\alpha}\partial_{t} +\sinh{\alpha}\partial_{\psi}, \label{over1}\\
\partial_{\psi} \rightarrow \cosh{\alpha}\partial_{\psi} +\sinh{\alpha}\partial_{t},\label{over2}\\
\partial_{a}-A_{a}\partial_{\psi} \rightarrow \partial_{a}-A_{a}\partial_{\psi} \label{over3}.
\end{gather}
%
Thus, locally the near-zone metric of the rotating black hole is just a boost of the near-zone Schwarzschild metric \cite{Emparan:2013xia}, although of course there is no globally defined coordinate transformation that relates them.

This local boost relation has momentous consequences for the large $D$ analysis of MP quasinormal modes. The most direct one is that the relation (\ref{over1}) seems to
imply that if we consider a perturbation of the MP black hole of the form $ e^{-i\omega t}e^{im\psi}$, then its frequency to leading order must be related to that of a perturbation of the Schwarzschild black hole $e^{-i\omega_{\text{Sch}} t}e^{im_\text{Sch}\psi}$
by
%
\begin{eqnarray}\label{trans}
\omega_{\text{Sch}} &=& \omega\cosh{\alpha} - m\sinh{\alpha} =\frac{\omega-a\, m}{\sqrt{1-a^2}}\,,\notag\\
m_{\text{Sch}} &=& -\omega\sinh{\alpha} + m\cosh{\alpha}=\frac{m-a\, \omega}{\sqrt{1-a^2}}\,.
\end{eqnarray}
%
In this case, Schwarzschild quasinormal frequencies would extend to MP ones as in eq.~\eqref{schboost}. However, as we discussed in the introduction, while this is true for some modes it does not hold for others, making the problem rather more subtle and interesting.

\section{Perturbations}
\label{sec:perts}

Cohomogeneity 1 of the solution directly implies the separability of variables in the metric perturbations $h_{\mu\nu}$ in any dimension. In order to make headway analytically it is highly desirable to also be able to decouple the perturbations, \ie\ reduce the problem to a set of decoupled second-order ordinary differential equations (analogous to the Teukolsky equation for Kerr). In general it is very difficult, if not impossible, to achieve this. However, the large $D$ expansion enables it through the boost relation described above.

\subsection{$\mathbb{CP}^N$ harmonic decomposition}
\label{subsec:cpnharm}

The general set up for perturbations of odd-$D$ equal-spin black holes has been described in \cite{Dias:2010eu} (see also \cite{Durkee:2010ea,Kunduri:2006qa}) so we will be brief. Schematically, we take 
\beq
h_{\mu\nu}= e^{-i\omega t}e^{i m\psi}\left[ f(r)\mathbb{Y}(x^a)\right]_{\mu\nu}\,,
\eeq
where the part of the perturbation depending on $r$ and on $\mathbb{CP}^N$ variables $x^a$ is classified according to the transformation properties under isometries of $\mathbb{CP}^{N}$. Scalar harmonics $\mathbb{Y}$ on $\mathbb{CP}^{N}$ are
defined by
%
\begin{eqnarray}\label{cpnscal}
\lp \mathcal{D}^{2}+\lambda  \rp\mathbb{Y}=0,
\end{eqnarray}
%
\beq\label{deflambda}
\lambda=\ell(\ell+2N)-m^{2}\,,
\eeq 
where $\mathcal{D}_{a}=\hat{\nabla}_{a}-imA_{a}$, with $\hat{\nabla}_{a}$ the covariant derivative with respect to $\hat{g}_{ab}$, and $\ell-|m|$ is an even non-negative number. 
By taking derivatives $\mathcal{D}_{a}$ of $\mathbb{Y}$ one can construct scalar-derived vector and tensor harmonics on $\mathbb{CP}^{N}$, which we denote as\footnote{The indices $\pm$ denote parity under the complex structure of $\mathbb{CP}^N$. Note also that for $\ell=m~(\ell=-m)$ modes, $\mathbb{Y}^{+}_{a}~(\mathbb{Y}^{-}_{a})$, $\mathbb{Y}^{++}_{ab}~(\mathbb{Y}^{--}_{ab})$ and $\mathbb{Y}^{+-}_{ab}$ vanish. 
For $\ell=m+2~(\ell=-m+2)$ modes, $\mathbb{Y}^{++}_{ab}~(\mathbb{Y}^{--}_{ab})$ vanishes.
}
%
\begin{eqnarray}\label{scderived}
\mathbb{Y}^{\pm}_{a}\,,\,~\mathbb{Y}^{\pm\pm}_{ab}\,,\,~\mathbb{Y}^{+-}_{ab}.
\end{eqnarray}
%
We consider gravitational pertubations that can be written in terms of these harmonics. These are called `scalar-type' perturbations.\footnote{$\mathbb{CP}^N$ tensor-type perturbations are stable \cite{Kunduri:2006qa}. Vector-type modes are expected to be stable.} 
We introduce the vielbein 
%
\begin{eqnarray}
e^{(0)}=\sqrt{\frac{G(r)}{H(r)}}\,dt\,,\quad e^{(1)}=\frac{dr}{\sqrt{G(r)}}\,,\quad e^{(2)}=r\sqrt{H(r)}\,(d\psi+A_{a}dx^{a}-\Omega(r)dt), 
\end{eqnarray}
%
and $e^{(i)}=r\hat{e}^{(i)}$ with $\hat{e}^{(i)}$ a vielbein on $\mathbb{CP}^{N}$. Then, following the notation of \cite{Dias:2010eu}, we write the scalar-type perturbations in the form
%
\begin{eqnarray}
&&
h_{\mu\nu}dx^{\mu}dx^{\nu} = e^{-i\omega t}e^{im\psi}\Bigl[ f_{AB}\mathbb{Y}e^{(A)}e^{(B)} 
+ 2r(f^{+}_{A}\mathbb{Y}^{+}_{a}+f^{-}_{A}\mathbb{Y}^{-}_{a})e^{(A)}dx^{a} \notag \\
&&~~~~~~~
+r^{2}\Bigl( H_{L}\mathbb{Y}\hat{g}_{ab} -\lambda^{-1/2}(H^{++}\mathbb{Y}^{++}_{ab}+H^{+-}\mathbb{Y}^{+-}_{ab}+H^{--}\mathbb{Y}^{--}_{ab}) \Bigr)
dx^{a}dx^{b} \Bigr], \label{hdef}
\end{eqnarray}
%
with $A,B={0,1,2}$. 

In appendix~\ref{app:sncpn} we explain the relation between $\mathbb{CP}^N$ harmonics and spherical harmonics of $S^{2N+1}$, and in particular how the eigenvalue parameter $\ell$ of a $\mathbb{CP}^N$ harmonic relates to the angular momentum number $\ell_\text{sph}$ of a $S^{2N+1}$ harmonic.
The conclusion is that the subspace of $\mathbb{CP}^N$ harmonics $\mathbb{Y}$, $\mathbb{Y}^\pm_a$ with parameter $\ell$ consists of scalars and scalar-derived vectors of $S^{2N+1}$ with 
\beq\label{scrule}
\ell_\text{sph}=\ell\qquad (S^{2N+1}\;\text{scalar-type)}
\eeq
and of $S^{2N+1}$ vectors with
\beq\label{vecrule}
\ell_\text{sph}=\ell\pm 1\qquad (S^{2N+1}\;\text{vector-type)}\,.
\eeq
We will use these relations to relate the limit $a=0$ of quasinormal modes of MP black holes to modes for the Schwarzschild black hole in the conventional representations of $SO(D-2)$. We will not be concerned with tensors of $S^{2N+1}$ since they do not have quasinormal frequencies at $\omega=\mc{O}(D^0)$.

\subsection{Decoupling the equations}
Eq.~\eqref{hdef} contains 16 perturbation functions, which satisfy a set of coupled ordinary differential equations \cite{Dias:2010eu}. In order to see how the large $D$ limit allows to decouple them, consider first the case $a=0$ of the Schwarzschild black hole. In the large $D$ limit the perturbation equations in the transverse-traceless (TT) gauge do
decouple if we take the perturbation variables to be
%
\begin{eqnarray} \label{SchD}
&f^{\text{(Sch)}}_{00}+f^{\text{(Sch)}}_{11},\qquad &f^{\text{(Sch)}}_{01},\qquad  f^{\text{(Sch)}}_{02}, \notag \\
&f^{\text{(Sch)}}_{00}-f^{\text{(Sch)}}_{11},\qquad  &f^{\text{(Sch)}}_{12},\qquad  f^{\text{(Sch)}}_{22}\,,\\
&
f^{\text{(Sch)}+}_{A}\,,\qquad &H^{\text{(Sch)}+-}. \notag
\end{eqnarray}
The other variables, $f_{A}^{\text{(Sch)}-}$, $H_{L}^{\text{(Sch)}}$ and $H^{\text{(Sch)}\pm\pm}$, are given by the TT gauge conditions. In this manner we can solve the perturbation equations order by order and find quasinormal modes of the Schwarzschild black hole at large $D$. 

The boost relationship between the near-zone geometries at large $D$ allows to decouple the perturbations in a similar way when rotation is present. After transforming by a boost the near-zone vielbeins, we obtain the perturbation variables at finite rotation in terms of the static ones (details in appendix~\ref{app:decoupl}). We denote the decoupling variables as 
\beq
F_{AB}(\sR)\,,\qquad F_{A}(\sR)\,,\qquad H^{+-}(\sR)\,,
\eeq
in terms of which the original variables are
%
\beqa\label{f00}
f_{00}=
\frac{\sR(F_{00}+F_{11})-2a (\sR-1 )F_{02} +a^2(\sR-1) F_{22}  }
{\sR -a^{2}(\sR-1)},
\eeqa
%
%
\beqa
f_{01} = \frac{\sR F_{01} -a(\sR-1)F_{12}}{\sqrt{\sR(\sR-a^{2}(\sR-1))}},
\eeqa
%
%
\beqa
f_{02} =\sqrt{\frac{\sR-1}{\sR}} 
\frac{(\sR+a^{2}(\sR-1))F_{02} -a\sR(F_{00}+F_{11} + F_{22})}{\sR-a^{2}(\sR-1)},
\eeqa
%
%
\beqa
f_{12} = \frac{\sqrt{\sR-1}(F_{12}-aF_{01})}{\sqrt{\sR-a^{2}(\sR-1)}},
\eeqa
%
%
\beqa
f_{22} =
\frac{\sR F_{22} +a^2(\sR-1)(F_{00}+F_{11})-2a(\sR-1)F_{02}}{\sR -a^{2}(\sR-1)},
\eeqa
%
%
\beqa
f_{0}^+ = \frac{\sR F_{0} -a(\sR-1)F_{2}}{\sqrt{\sR(\sR-a^{2}(\sR-1))}},
\eeqa
%
%
\beqa
f_{2}^+ = \frac{\sqrt{\sR-1}(F_{2}-aF_{0})}{\sqrt{\sR-a^{2}(\sR-1)}},
\eeqa
%
%
\beqa\label{f11}
f_{11} =F_{00}-F_{11}\,,\qquad
f^{+}_{1}=F_{1}\,.
\eeqa
%
The remaining variables, $f^{-}_{A}$, $H_{L}$ and $H^{\pm\pm}$ are again determined by the TT gauge conditions. 

If we 
expand the decoupling variables at large $N$
\begin{eqnarray}
F_{AB}(\sR)=\sum_{k\geq 0}\frac{F_{AB}^{(k)}(\sR)}{N^{k}},\quad
F_{A}(\sR)=\sum_{k\geq 0}\frac{F_{A}^{(k)}(\sR)}{N^{k-1/2}},\quad
H^{+-}(\sR)=\sum_{k\geq 0}\frac{H^{+-(k)}(\sR)}{N^{k-1}}\,,
\end{eqnarray}
%
(the difference in the scaling of $N$ comes from the normalization of harmonics), then the perturbation equations in TT gauge derived in \cite{Dias:2010eu} can also be expanded, and at the $k$-th order take the form
%
\begin{eqnarray}\label{F00eq}
\sR^{2}(\sR-1)F_{00}^{(k)''}(\sR)+\sR(2\sR-1)F_{00}^{(k)'}(\sR)+F_{00}^{(k)}(\sR) = \mathcal{S}_{F_{00}}^{(k)},
\end{eqnarray}
%
%
\begin{eqnarray}
\sR(\sR-1)^{2}F_{01}^{(k)''}(\sR)+(\sR-1)(2\sR-1)F_{01}^{(k)'}(\sR)-F_{01}^{(k)}(\sR) = \mathcal{S}_{F_{01}}^{(k)},
\end{eqnarray}
%
%
\begin{eqnarray}
(\sR-1)F_{02}^{(k)''}(\sR)+2F_{02}^{(k)'}(\sR) = \mathcal{S}_{F_{02}}^{(k)},
\end{eqnarray}
%
%
\begin{eqnarray}
\sR(\sR-1)^{2}F_{11}^{(k)''}(\sR)+(\sR-1)(2\sR-1)F_{11}^{(k)'}(\sR)-F_{11}^{(k)}(\sR) = \mathcal{S}_{F_{11}}^{(k)},
\end{eqnarray}
%
%
\begin{eqnarray}
(\sR-1)F_{12}^{(k)''}(\sR)+2 F_{12}^{(k)'}(\sR) = \mathcal{S}_{F_{12}}^{(k)},
\end{eqnarray}
%
%
\begin{eqnarray}
(\sR-1)F_{22}^{(k)''}(\sR)+(2\sR-1)F_{22}^{(k)'}(\sR) = \mathcal{S}_{F_{22}}^{(k)},
\end{eqnarray}
%
%
\begin{eqnarray}
\sR(\sR-1)^{2}F_{0}^{(k)''}(\sR)+(\sR-1)(2\sR-1)F_{0}^{(k)'}(\sR)-F_{0}^{(k)}(\sR) = \mathcal{S}_{F_{0}}^{(k)},
\end{eqnarray}
%
%
\begin{eqnarray}
4\sR^{2}(\sR-1)^{2}F_{1}^{(k)''}(\sR)+4\sR(\sR-1)(2\sR-1)F_{1}^{(k)'}(\sR)-F_{1}^{(k)}(\sR) = \mathcal{S}_{F_{1}}^{(k)},
\end{eqnarray}
%
%
\begin{eqnarray}
(\sR-1)F_{2}^{(k)''}(\sR)+2 F_{2}^{(k)'}(\sR) = \mathcal{S}_{F_{2}}^{(k)},
\end{eqnarray}
%
%
\begin{eqnarray}\label{Hpmeq}
\sR(\sR-1)H^{+-(k)''}(\sR)+ (2\sR-1)H^{+-(k)'}(\sR) = \mathcal{S}_{H^{+-}}^{(k)}\,,
\end{eqnarray}
%
where $\mathcal{S}^{(k)}$ are the source terms at order $k$, which are built out of the lower-order solutions in the conventional manner. We omit their form, which is lengthy but straightforward to derive. 

Two important remarks are in order. First, although strictly speaking the decoupling of equations occurs only at the leading large $D$ order, this is enough for the purpose of solving the equations perturbatively in the large $D$ expansion. All the coupling between variables resides in the source terms $\mathcal{S}^{(k)}$, and therefore the decoupled structure of the leading order equations persists to all higher orders. Second, since the source terms beyond the leading order are not obtained by simply boosting the ones for Schwarzschild, the higher-order \textit{solutions} will not preserve this boost property.

In other words, the boost relation \eqref{transL2} is on the one hand crucial for the solvability of the system at finite rotation, but on the other hand it does not trivialize entirely its solution and allows to capture dynamics specific to the rotation of the black holes.

\section{Quasinormal mode analysis}
\label{sec:qnman}

The solution of equations (\ref{F00eq}--\ref{Hpmeq}) is straightforward to obtain once the boundary conditions are specified.
Among all perturbations, quasinormal modes are characterized by being outgoing at asymptotically flat infinity, and ingoing at the future horizon.

\paragraph{Asymptotic boundary conditions.} The asymptotically flat region is in the far-zone where the metric is exactly flat to leading order and the general solution is of the form
\beq
h_{\mu\nu}\sim r^{-D/2}\lp A^\text{(out)}H_\nu^{(1)}(\omega r)+ A^\text{(in)}H_\nu^{(2)}(\omega r)\rp
\eeq
where $H_\nu$ are Hankel functions with index $\nu\propto D$. Purely outgoing modes are obtained when $A^\text{(in)}=0$. Generically, this results into a restriction on the behavior of the field in the overlap zone, which then provides a boundary condition for the near-zone solutions. 

For frequencies $\omega=\mc{O}(D)$ the condition $A^\text{(in)}=0$ results in an important constraint on the overlap-zone field. However, it is much less restrictive for the frequencies $\omega=\mc{O}(D^0)$ that we study here. For these, we have
\beq
r^{-D/2}H_\nu^{(1,2)}(\omega r)=\mc{O}(D^{D/2})\times \sR^{-1}
\eeq
and the difference between the two solutions is suppressed by factors $\mc{O}(D^{-D})$, which are not visible in the $1/D$ expansion. This implies that for $\omega=\mc{O}(D^0)$, as long as the near-zone solution satisfies at large $\sR$ the boundary condition
%
\beq\label{asymbc}
h_{\mu\nu} = \mc{O}(\sR^{-1}),
\eeq
%
then it can always be extended into an outgoing solution in the asymptotically flat region. Thus  \eqref{asymbc} is the condition to impose on our near-zone modes.

\paragraph{Horizon boundary conditions.} The regularity conditions at the horizon
can be easily obtained by changing to ingoing Eddington-Finkelstein coordinates \cite{Dias:2010eu}. It requires that near $\sR=1$ the perturbations behave in the form
%
\beqa
(\sR -1)f_{00},\ f_{01}-f_{00},\ \frac{f_{00}-2f_{01}+f_{11}}{\sR -1},\ \sqrt{\sR -1}f_{02}, &&\notag\\ 
 \frac{f_{12}-f_{02}}{\sqrt{\sR -1}},\ f_{22},\  
f_{2}^{\pm},\ \sqrt{\sR -1}f_{0}^{\pm},\ \frac{f_{0}^{\pm}-f_{1}^{\pm}}{\sqrt{\sR -1}},\  H^{\pm\pm},\ H_{L} &&\propto (\sR -1)^{-2i(\omega- m\Omega_{H})/\kappa} \,, 
\eeqa
where $\kappa$ is the surface gravity at the horizon \eqref{surfg}. Expanding at large $N$ one gets\footnote{We can see here that close to extremality, where $\kappa\to 0$, the next-to-leading order corrections are expected to be $\sim (\ln N)/N$.}
%
\begin{eqnarray}
&&
(\sR -1)^{-2i(\omega- m\Omega_{H})/\kappa} 
=
1 -\frac{i(\omega -a m)}{2\sqrt{1-a^{2}}N} \log{(\sR-1)}-\frac{(\omega-am)^{2}}{8(1-a^{2})N^{2}}(\log{(\sR-1)})^{2}\notag \\
&&\qquad
+\frac{i\log{(\sR-1)}}{4(1-a^{2})^{3/2}N^{2}}\Bigl[ 
-2a^{2}( \omega -am )-( (1-2a^{2})\omega +am )\log{(1-a^{2})} 
\Bigr]  \notag \\
 &&\qquad +\mc{O}(N^{-3})\,.
\end{eqnarray}
The second and third terms in the r.h.s.\ are clearly the boosted form, under \eqref{trans}, of the expressions for $a=0$. However, the fourth term, which only enters at next-to-next-to-leading order, vanishes when $a=0$ and hence does not follow in that manner. This is another reason why those quasinormal frequencies that are determined by boundary conditions at this order cannot be obtained by a boost of Schwarzschild modes.

\medskip

We proceed to succintly present the results from the solution to the equations, omitting the goriest details. Some arguments that follow would break down for $\ell=0$ modes, but in appendix~\ref{app:ell0} we prove their stability for any $N$. Thus in the following we take $\ell>0$.

\paragraph{Leading order.}
The solution that satisfies the boundary conditions is
%
\begin{alignat}{3}
&F_{02}^{(0)}(\sR) = \frac{A_{0}}{\sR-1}\,, &&\qquad F^{(0)}_{22}(\sR) = 0\,, &&\qquad F^{(0)}_{00}(\sR) = \frac{C_{0}}{\sR}\,,\notag \\
&F^{(0)}_{11}(\sR) = \frac{D_{0}}{\sR-1}\,, &&\qquad F^{(0)}_{01}(\sR) = \frac{D_{0}}{\sR-1}\,, &&\qquad F^{(0)}_{12}(\sR) = \frac{A_{0}}{\sR-1}, \\
&F^{(0)}_{0}(\sR) =  \frac{G_{0}}{\sqrt{\sR(\sR-1)}}\,, &&\qquad F^{(0)}_{2}(\sR) =0\,, &&\qquad F^{(0)}_{1} = \frac{G_{0}}{\sqrt{\sR(\sR-1)}} , \notag \\
&H^{+-(0)}(\sR) =0.&&&\notag
\label{LOsol}
\end{alignat}
$A_{0}, C_{0}, D_{0}$ and $G_{0}$ are integration constants that at this order remain undetermined by boundary conditions.
The ingoing condition becomes non-trivial only at sub-leading order, where it picks a specific combination among $A_{0}, C_{0}, D_{0}$ and $G_{0}$.

\paragraph{Next-to-leading order.}

At this order we obtain
%
\begin{gather}
G_{0}=0, \\
D_{0}=-2a A_{0}+C_{0}
\left( -i(\omega-am)\sqrt{1-a^{2}} +\ell-1-a^{2}\ell \right), \label{CDeq}
\end{gather}
%
and either $A_{0}=0$ or $\omega = \omega_{\ell,m}^{(1)}$
%
with
\beq
\omega_{\ell,m}^{(1)} = a m -i \ell\sqrt{1-a^{2}}\,. \label{qnm1}
\eeq
Thus, for a perturbation with $A_{0}\neq 0$ we obtain quasinormal modes with the frequency $\omega_{\ell,m}^{(1)}$. Since they always have $\text{Im}\,\omega <0$, these modes are stable. In sec.~\ref{subsec:static} below we show that this frequency is a boost of a Schwarzschild quasinormal frequency.

It would seem that we could obtain another quasinormal mode frequency if in (\ref{CDeq}) we set $A_{0}=D_{0}=0$. However it turns out that this choice is not valid since it does not satisfy the horizon regularity condition at the next order.

\paragraph{Next-to-next-to-leading order.}

We set $A_{0}=0$ in order to obtain quasinormal modes different than the ones found at the previous order. 

We find one solution for regular perturbations with frequency $\omega =am -i \ell\sqrt{1-a^2}$ (the same as \eqref{qnm1}). However, we show in appendix~\ref{app:gauge} that this is a gauge mode. 
Physical regular perturbations, instead, must have frequencies satisfying
%
\begin{eqnarray}
0=&&
\frac{1}{\omega -a(m+2)+i(\ell-2)\sqrt{1-a^2}} 
\Bigl[\,\omega ^3 
+\,\omega ^2 \left(-3 a m+i\left(3 \ell-4 \right)\sqrt{1-a^2} \right) \notag \\
&&\quad
+\,\omega  \left(3 a^2 \ell^2-6 i am \sqrt{1-a^2}\left(\ell - 1\right) -6a^2 \ell+3 a^2 m^2-3 \ell^2+7 \ell-4\right) \notag \\
&&\quad
+\,a m\lp 2+(4 a^2-5)\ell+3(1-a^2)\ell^2-a^2 m^2\rp \notag \\
&&\quad
+\,i \sqrt{1-a^2} \left(-\left(1-a^2\right) \ell^3+\left(3-2a^2\right) \ell^2+\ell \left(3 a^2 m^2-2\right)-2 a^2 m^2\right)
\Bigr]\,. \label{QNMeq}
\end{eqnarray}
%
This cubic equation is one of the main results of this article:  its roots yield the frequencies $\omega_{\ell,m}$ of three independent quasinormal modes to leading order at large $D$.\footnote{We keep the denominator in \eqref{QNMeq} to cancel a spurious root of the numerator when $\ell=m$, see below.}\ Note that under $m\to -m$ we have $\text{Re}\,\omega\to -\text{Re}\,\omega$ and $\text{Im}\,\omega\to \text{Im}\,\omega$. In the following we only consider $m\geq 0$.

We have also obtained the next-order correction to \eqref{qnm1} and \eqref{QNMeq}, which determine the $1/N$ corrections to the quasinormal frequencies. This result is much too lengthy to show explicitly here, and so is instead given in the \textit{Mathematica} file attached to this article. For our purposes it will suffice that, after we extract the main conclusions of the leading-order results, we discuss how the next-to-leading order corrections modify them.

\section{Instabilities}
\label{sec:instab}

Our analysis yields in general four quasinormal mode frequencies. Besides \eqref{qnm1}, the other three frequencies are roots of a cubic equation which is straightforward to solve in analytic form (\eg\ using the del Ferro-Tartaglia-Cardano solution). Then we can easily plot $\omega_{\ell,m}$ as a function of $a$. However, the expressions are not very illuminating, so we will not present them here. The following analysis extracts the most important properties concerning instability.

\subsection{Non-axisymmetric modes}
\label{sec:naxi}

The solutions do take a simple form when $\ell=m$. In this case  eq.~(\ref{QNMeq}) gives only two roots, which together with \eqref{qnm1} yield\footnote{The absence of a fourth mode is consistent with the point noted at the end of appendix~\ref{app:sncpn}.}
%
\begin{align}
\omega_{m,m}^{(1)}&=am -i m\sqrt{1-a^2}\,,\label{qnmlm1}  \\
\omega_{m,m}^{(2)}&=a(m-1)-\sqrt{(m-1)(1-a^{2})}-i\bigl[ (m-1)\sqrt{1-a^{2}} +a\sqrt{m-1} \,\bigr]\,,\label{qnmlm2}\\
\omega_{m,m}^{(3)}&=a(m-1)+\sqrt{(m-1)(1-a^{2})}-i\bigl[ (m-1)\sqrt{1-a^{2}} -a\sqrt{m-1}\,\bigr]\,\label{qnmlm3}
\end{align}
%
%
\begin{figure}[t]
 \begin{center}
  \includegraphics[width=.47\textwidth,angle=0]{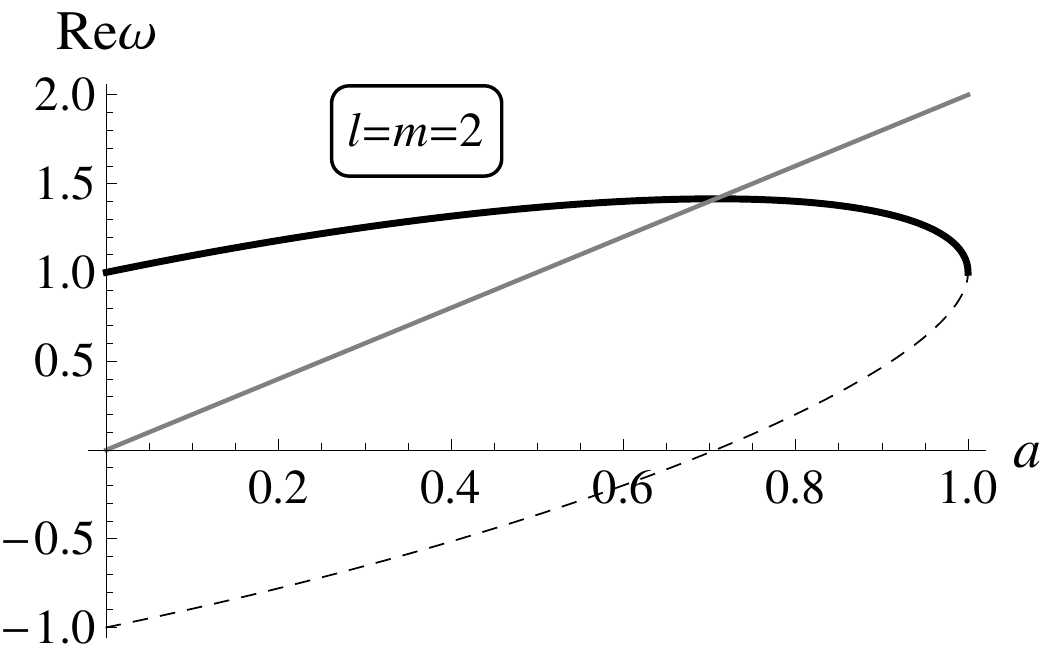}
  \hspace{5mm}
  \includegraphics[width=.47\textwidth,angle=0]{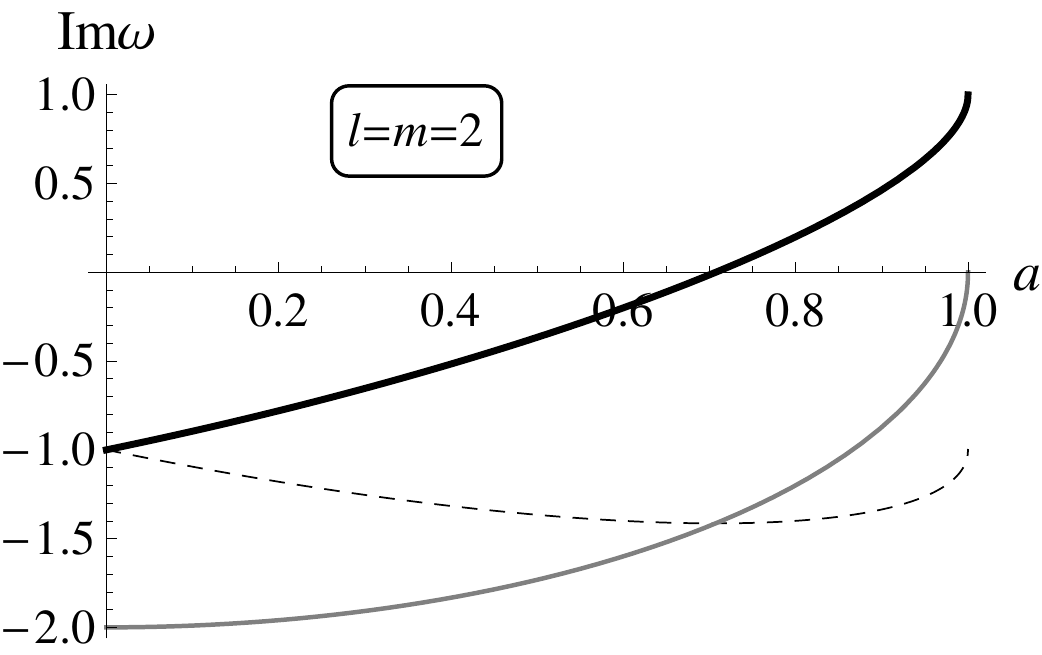}
   \end{center}
 \vspace{-5mm}
 \caption{\small Real and imaginary frequencies of quasinormal modes $(\ell,m)=(2,2)$ (in units $r_+=1$). Here  and in the next plots the thick black line is the mode $\omega_{\ell,m}^{(3)}$ that becomes unstable, the thin gray line is $\omega_{\ell,m}^{(1)}$ in \eqref{qnm1}, and the discontinuous lines are the other quasinormal modes from \eqref{QNMeq}. For $\ell=2$ the instability occurs at $a_c=1/\sqrt{2}$. Also at that value the black line $\text{Re}\,\omega$ cuts the gray line $\omega=am$ that marks the superradiant bound. Higher $\ell=m$ modes show the same qualitative behavior. Note that $\text{Re}\,\omega_{m,m}^{(3)}\neq 0$ at $a=0$. The large $N$ expansion breaks down near the extremal limit, so the results very close to $a=1$ become less reliable.}
 \label{fig:lm2}
\end{figure}
%
(see fig.~\ref{fig:lm2}). 
The mode $\omega_{m,m}^{(3)}$ shows a dynamical instability, $\text{Im}\,\omega>0$, for
%
\begin{eqnarray}
a> \sqrt{1-\frac{1}{m}}\,. \label{acrilm}
\end{eqnarray}
%
The onset of the instability, where $\text{Im}\,\omega=0$, coincides with the threshold at which the mode becomes superradiant, $\omega=a m$. For $a>a_c$ this mode has $\text{Re}\,\omega <a m$.
In appendix~\ref{app:gauge} we show that the apparent zero mode at $\ell=m=1$ is a gauge mode.

\begin{figure}[t]
 \begin{center}
  \includegraphics[width=.47\textwidth,angle=0]{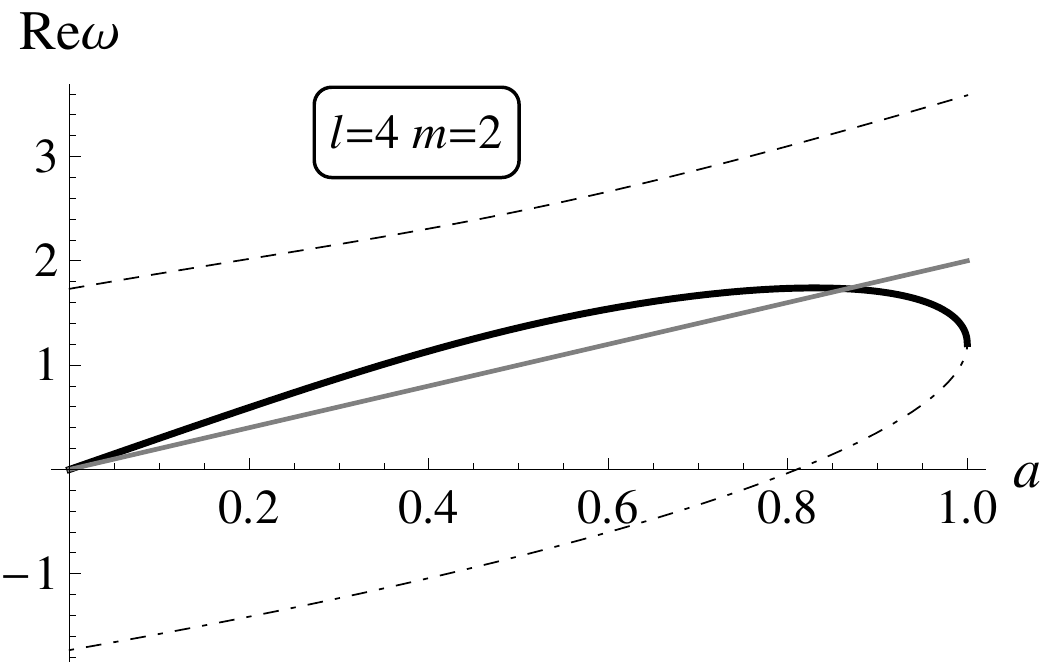}
  \hspace{5mm}
  \includegraphics[width=.47\textwidth,angle=0]{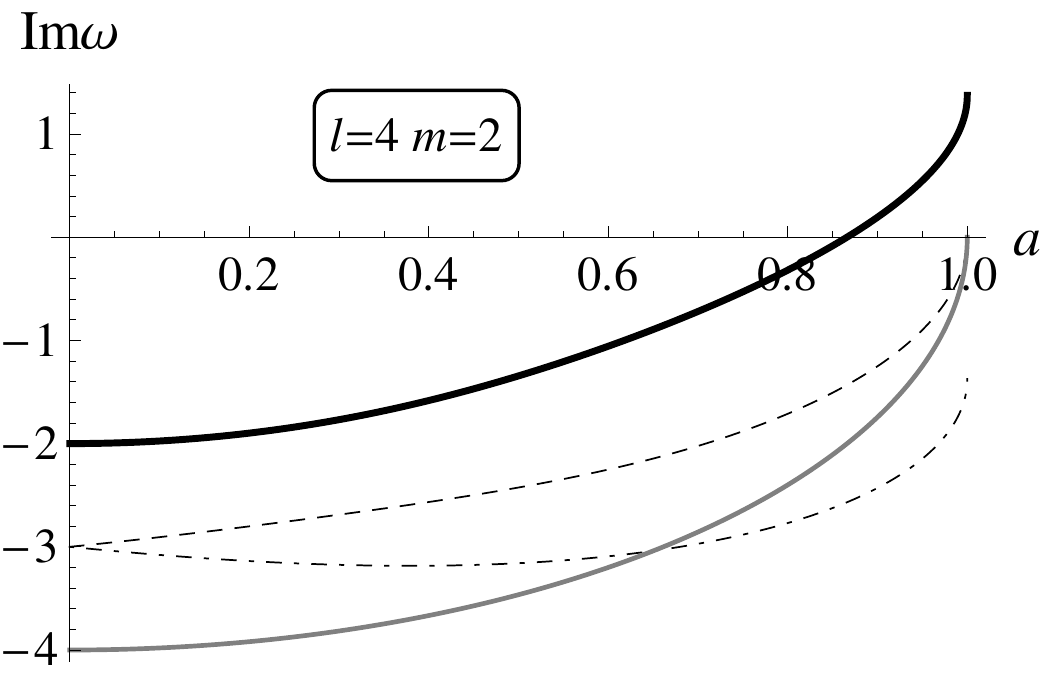}
   \end{center}
 \vspace{-5mm}
 \caption{\small Real and imaginary frequencies of quasinormal modes $(\ell,m)=(4,2)$. The  instability sets in at $a_c=\sqrt{3}/2$, where the unstable mode is also at the superradiant threshold $\text{Re}\,\omega=a m$. At $a=0$ this mode becomes purely imaginary.}
 \label{fig:l4m2}
\end{figure}

For $\ell\neq m$ the solutions take a more complicated form but we can easily plot the  three roots $\omega_{\ell,m}$ of the cubic in \eqref{QNMeq} as functions of $a$, to see that the imaginary part of one of them changes from negative to positive at a critical value, while the other two modes remain stable (see fig.~\ref{fig:l4m2}). Indeed, it is straightforward to check that if
\beq
a=a_c= \sqrt{1-\frac{1}{\ell}}, \label{acrigen}
\eeq
then \eqref{QNMeq} is solved, for any $m$, by
\beq
\omega =a_c m\,,
\eeq
\ie\ there is a mode that becomes purely real at $a=a_c$, with a frequency that sits precisely at the threshold for superradiance. If we now consider
\beq
a=a_c+\delta a
\eeq
with small $\delta a$, then we find the solution
\beq\label{nearc}
\omega_{\ell,m}= a_c m \lp 1-\frac{\ell}{\ell^2(\ell-1)+m^2}\delta a\rp
+i\frac{\ell^2(\ell-2)+m^2}{\sqrt{\ell}\lp\ell^2(\ell-1)+m^2\rp}\delta a+\mc{O}\lp\delta a^2\rp\,,
\eeq
so when $\delta a>0$ the mode becomes unstable, and is within the superradiant range $\text{Re}\,\omega_{\ell,m}<a m$. 

As $a$ increases from zero, the first mode to reach $a_c$ and become unstable is $\ell=m=2$, hence the bound \eqref{naxibound}. Equation~\eqref{nearc} also shows that for fixed $\ell$, the fastest growing, most unstable mode, near the critical point is the one with the largest $m$, namely $m=\ell$. This continues to be true up to $a=1$. 

Although modes with lower $\ell$ become unstable at lower rotations, those with higher $\ell$ have larger $\text{Im}\, \omega$ at sufficiently high rotation. Therefore, at very high rotations the instability is dominated by modes with high $\ell=m$, \ie\ highly inhomogeneous modes. 

Observe that $a_c$ in \eqref{acrigen} depends on $\ell$ but not on $m$. Also, the critical rotation for the dominant mode $\ell=m=2$ coincides with the ultraspinning bound $a_\text{ultra}=1/\sqrt{2}$. These two features will cease to hold when $1/N$ corrections are included.

\subsection{Axisymmetric modes}
\label{sec:axisym}

When $m=0$, eq.~\eqref{QNMeq} becomes of the form 
\beq
\omega^3+i\, c_2\,\omega^2+c_1\,\omega+i\, c_0=0
\eeq
with real coefficients $c_i$ for all $a\in [0,1]$. Hence one of its roots, $\omega^{(0)}$, is purely imaginary, while the other two are related by $\text{Re}\,\omega^{(+)}=-\text{Re}\,\omega^{(-)}$ and $\text{Im}\,\omega^{(+)}=\text{Im}\,\omega^{(-)}$.

Besides the quasinormal frequency $\omega_{\ell,0}^{(1)}$ in \eqref{qnm1}, it is possible to find the other three in explicit form, but again the result is not quite transparent. The exception are the lowest modes, $\ell=2$, with frequencies 
%
\begin{eqnarray}
\omega_{2,0}^{(0)}=0\,,\qquad\omega_{2,0}^{(\pm)}=\pm\sqrt{1+a^{2}}-i\sqrt{1-a^{2}}\,.
\end{eqnarray}
%
The zero mode $\omega_{2,0}^{(0)}$ is a stationary perturbation that corresponds to a variation in the angular 
momentum of the MP black hole along the family of stationary solutions. It does not signal a dynamical instability, nor a branching into a new family of black holes. The other $(\ell,m)=(2,0)$ modes are always stable.

For $\ell\geq 4$ the mode $\omega^{(0)}_{\ell,0}$ with purely imaginary frequency (but not the others) changes sign and becomes unstable at the same critical value \eqref{acrigen}. For $a=a_c+\delta a$ its frequency is given by eq.~\eqref{nearc} with $m=0$, confirming that the mode becomes unstable for $a>a_c$. At $a=a_c$ this is a zero mode, and we expect that for each value of $\ell\geq 4$ this gives rise to a new branch of solutions. Such zero modes have been found for equal-spin odd-$D$ black holes in \cite{Dias:2010eu}.
\begin{figure}[t]
 \begin{center}
  \includegraphics[width=.47\textwidth,angle=0]{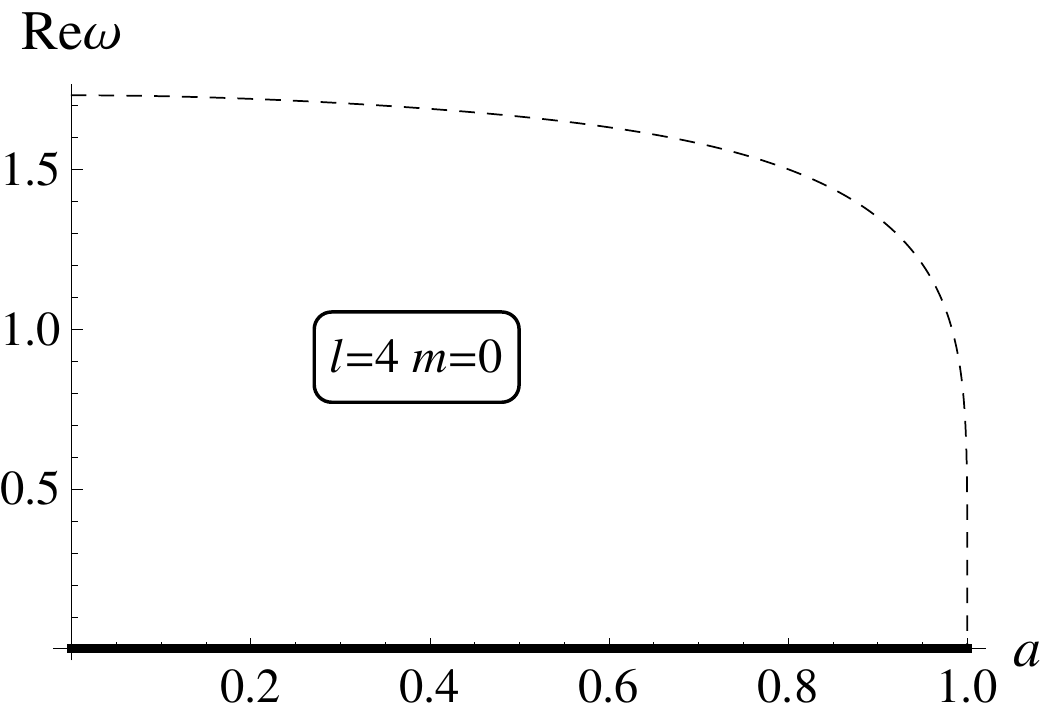}
  \hspace{5mm}
  \includegraphics[width=.47\textwidth,angle=0]{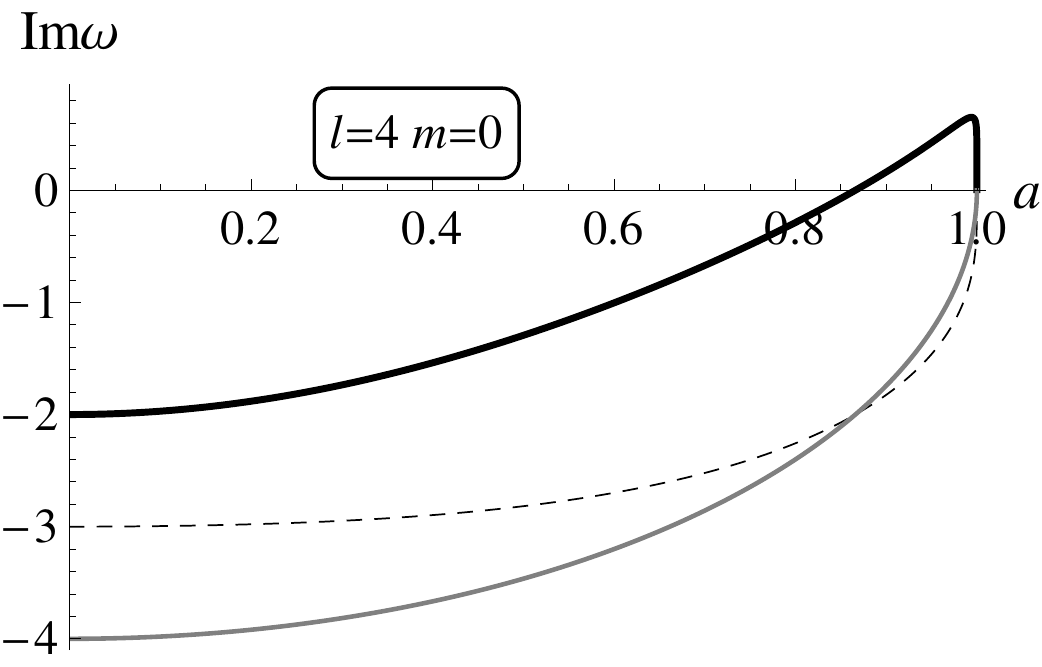}
   \end{center}
 \vspace{-5mm}
 \caption{\small Real and imaginary frequencies of quasinormal modes $(\ell,m)=(4,0)$. The thick solid line is the purely imaginary mode $\omega^{(0)}_{\ell,0}$ that becomes unstable at $a_c=\sqrt{3}/2$. The gray-line mode $\omega_{\ell,0}^{(1)}$ is also purely imaginary, but stable. The dashed line is $\omega_{\ell,0}^{(+)}$, and we omit $\omega_{\ell,0}^{(-)}=-\lp\omega_{\ell,0}^{(+)}\rp^*$.}
 \label{fig:l4m0}
\end{figure}
The first axisymmetric instability to appear has $\ell=4$, thus giving the bound \eqref{axibound}. In fig.~\ref{fig:l4m0} we show the stable and unstable modes for $\ell=4$.

\subsection{Limit $a=0$ and Schwarzschild quasinormal modes}
\label{subsec:static}

When $a=0$ our results must yield quasinormal frequencies of Schwarzschild black holes. In calculations that will be presented elsewhere we have computed this same spectrum following the Kodama-Ishibashi analysis \cite{Kodama:2003jz} in which the gravitational perturbations are classified into scalar-type, vector-type and tensor-type modes of $S^{D-2}$ with angular momentum number $\ell_{\text{sph}}$. We have found that the gravitational scalar-type and vector-type perturbations, but not the tensors, have modes with frequency $\omega=\mc{O}(D^0)$ of the form
\beq\label{Schs}
\omega^{\text{scalar}}_{(\pm)} = \pm\sqrt{\ell_{\text{sph}}-1}-i(\ell_{\text{sph}}-1)\,,
\eeq
and
%
\begin{eqnarray}\label{Schv}
\omega^{\text{vector}} = -i(\ell_{\text{sph}}-1)\,.
\end{eqnarray}
The scalar-type modes with $\ell_{\text{sph}}=0$ are actually variations of the black hole mass, and $\ell_{\text{sph}}=1$ is a gauge mode \cite{Kodama:2003jz}. The vector-type mode $\ell_{\text{sph}}=1$ only adds angular momentum to the black hole.  
We can now use the translation rules \eqref{scrule} and \eqref{vecrule} in order to relate these modes to the limit $a=0$ of the MP modes. 

The identification is simpler for modes with $\ell=m$, since in the spherically-symmetric case the frequency does not depend on $m$. Taking $a=0$ in \eqref{qnmlm1}, \eqref{qnmlm2}, \eqref{qnmlm3} suggests that we identify $\omega_{\ell,\ell}^{(1)}$ with a vector-type mode with $\ell_{\text{sph}}=\ell+1$ and $\omega_{\ell,\ell}^{(2),(3)}$ with the two scalar-type modes with $\ell_{\text{sph}}=\ell$. As mentioned at the end of app.~\ref{app:gauge}, when $\ell=m$ the $S^{D-2}$-vector with $\ell_{\text{sph}}=\ell-1$ is absent, hence the three modes of the MP black hole are in one-to-one correspondence with the three Schwarzschild modes \eqref{Schs} and \eqref{Schv}. However, when $\ell\neq m$, two different MP modes with the same $\ell$ must connect as $a\to 0$ to purely imaginary vector modes \eqref{Schv} with  $\ell_{\text{sph}}=\ell\pm 1$. Hence when $\ell\neq m$ there must be two modes that become purely imaginary as $a\to 0$ with $\omega_{\ell,m}\to -i\ell,\,-i(\ell-2)$.

Let us first consider the perturbation mode (\ref{qnm1}) obtained when $A_{0}\neq 0$. When $a=0$ the parameter $A_{0}$ generates a perturbation $F_{02}\propto f_{02}^\text{(Schw)}\propto g_{t\psi}$. So this mode originates at $a=0$ as a vector-type perturbation of Schwarzschild. Indeed, the frequency $\omega_{\ell,m}^{(1)}$ in \eqref{qnm1} is the result of boosting in the form \eqref{schboost} the Schwarzschild frequency $\omega^{\text{vector}}$ with $\ell_{\text{sph}}=\ell+1$. 

We can identify the other modes by studying the solutions of eq.~(\ref{QNMeq}) at $a=0$, and by introducing the rotation $a$ as a $\mc{O}(N^{-1})$ small perturbation on the Schwarzschild solution. 
The three MP modes that we obtain when $A_{0}=0$ can then be seen to originate at $a=0$ from
the two scalar-type modes with $\ell_{\text{sph}}=\ell$ and the vector-type mode with $\ell_{\text{sph}}=\ell- 1$. These MP modes are not simply boosts of Schwarzschild modes. 

For illustration, in figs.~\ref{fig:l4m2} and \ref{fig:l4m0} the black and gray lines connect at $a=0$ to Schwarzschild vector-type modes \eqref{Schv} with $\ell_\text{sph}=4\pm 1$, and the discontinuous lines to scalar-type modes \eqref{Schs} with $\ell_\text{sph}=4$.

It is interesting to observe that the origin at $a=0$ of the modes that become unstable for a given $\ell$ is different --- vector- or scalar-type --- depending on the value of $m$.  The unstable mode frequencies with $\ell=m$ become, when $a\to 0$,
\beq
\omega_{m,m}^{(3)}\rightarrow \sqrt{\ell-1}-i(\ell-1)
\eeq
corresponding to $\omega^{\text{scalar}}_{(+)}$ above. In fact all the unstable modes with sufficiently high $m$ come from scalar-type perturbations of Schwarzschild with $\text{Re}\,\omega\neq 0$.\footnote{The specific value of $m$ that defines ``sufficiently high $m$'' depends on $\ell$.}
In contrast, the unstable axisymmetric perturbations $m=0$ (and also those with low enough $m$) originate from Schwarzschild vector-type perturbations: when $a\to 0$ their purely imaginary frequency 
\beq
\omega\to -i(\ell-2)
\eeq
matches \eqref{Schv} with $\ell=\ell_{\text{sph}}+1$. 


\subsection{Instability to next-to-leading order}

The extension of the perturbation analysis to the next order in the $1/N$ expansion is computationally time-consuming but otherwise straightforward.

In the units $r_0=1$ that we are using we have that
\beq
r_+=1+\frac{\ln\lp 1-a^2\rp}{2N} +\mc{O}\lp N^{-2}\rp\,,
\eeq
the ultraspinning bound on the rotation is
\beq
a_\mathrm{ultra}=\frac{1}{\sqrt{2}}\lp 1-\frac{\ln 2}{2N}\rp +\mc{O}\lp N^{-2}\rp\,,
\eeq
and the frequency at the superradiant threshold is
\beq\label{supradb}
\omega=am\lp1 -\frac{\ln (1-a^2)}{N}\rp +\mc{O}\lp N^{-2}\rp\,.
\eeq

The quasinormal frequencies are expanded like
\beq
\omega_{\ell,m}=\omega_{\ell,m}|_0+\frac{\omega_{\ell,m}|_1}{N} +\mc{O}\lp N^{-2}\rp\,.
\eeq
The equation that determines $\omega_{\ell,m}|_1$ can be found in the \textit{Mathematica} file attached. We shall only quote the result for the mode $\omega^{(1)}_{m,m}$, for which  $\omega^{(1)}_{m,m}|_0$ is given by \eqref{qnmlm1}, since its correction becomes simple enough,
\beqa
\omega^{(1)}_{m,m}|_1&=&
\frac{m}{2 \left(a^2-1\right)}\Bigl[
2 a^5 m+a^3 (2-4 m)+i  a^2 (3 m-2)\sqrt{1-a^2}-i m\sqrt{1-a^2} \notag \\
&&
-2 i  a^4 m\sqrt{1-a^2}+\left(2 a^3-2 i a^2\sqrt{1-a^2}
   +i \sqrt{1-a^2}-2 a\right) \ln \left(1-a^2\right) \notag \\
&&
+2 a(m-1)
\Bigr]\,.
\end{eqnarray} 
%

Although the expression for other modes is very complicated, we can nevertheless identify that the instability threshold for a mode $(\ell,m)$ lies at
\beq\label{acnlo}
a_c=\sqrt{1-\frac1{\ell}}\lp 1-\frac{1}{2 N}\lp \frac{m^2}{2\ell^2}+\ln\ell\rp\rp +\mathcal{O}\lp N^{-2}\rp\,.
\eeq
When the critical rotation is measured in units of $r_+$ instead of $r_0$, we obtain \eqref{crita}. In order to convert this result to units of $a_{\text{ext}}$, we use
\beq\label{aextrp}
\frac{a_\text{ext}}{r_+}=\sqrt{\frac{N}{N+1}}=1-\frac1{2N}+\mathcal{O}\lp N^{-2}\rp
\eeq
so that
\beq\label{acaext}
\frac{a_c}{a_\text{ext}}=\sqrt{1-\frac1{\ell}}\lp 1+\frac{1}{2 N}\lp 1-\frac{m^2}{2\ell^2}\rp\rp +\mathcal{O}\lp N^{-2}\rp\,.
\eeq
The $\ell=m=2$ bar-mode instability is then present for
\beq
\frac{a}{a_\text{ext}}>\frac1{\sqrt{2}}\lp 1+\frac1{4N}\rp+\mathcal{O}\lp N^{-2}\rp\,,
\eeq
and therefore occurs at lower rotation than the onset of ultraspinning regime
\beq
\frac{a_\text{ultra}}{a_\text{ext}}=\sqrt{\frac{N+1}{2N}}=\frac1{\sqrt{2}}\lp 1+\frac1{2N}\rp+\mathcal{O}\lp N^{-2}\rp\,.
\eeq
Axisymmetric instabilities instead are present only at rotations
\beq
\frac{a}{a_\text{ext}}>\frac{\sqrt{3}}{2}\lp 1+\frac1{2N}\rp+\mathcal{O}\lp N^{-2}\rp
\eeq
entirely within the ultraspinning regime.

One can also check that at the critical rotation the real frequency takes the value \eqref{supradb} and therefore lies at the threshold of superradiance. Furthermore, the axisymmetric modes $\ell\geq 4$ that become unstable are purely imaginary.

%
\begin{figure}[t]
 \begin{center}
  \includegraphics[width=.47\textwidth,angle=0]{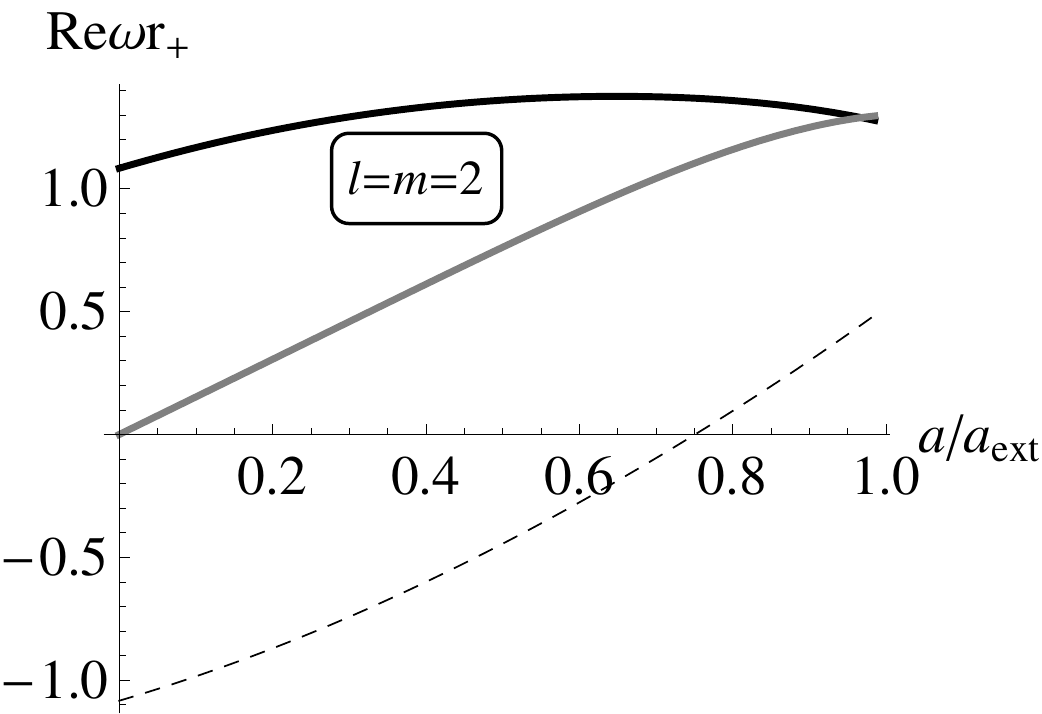}
  \hspace{5mm}
  \includegraphics[width=.47\textwidth,angle=0]{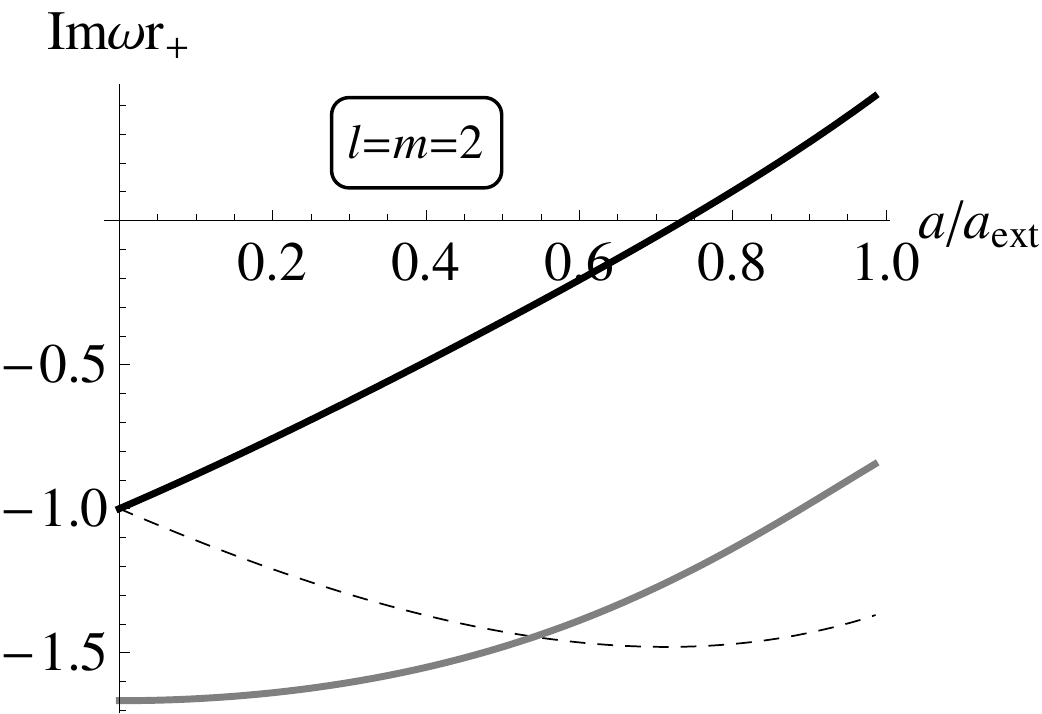}
   \end{center}
 \vspace{-5mm}
 \caption{\small Same as fig.~\ref{fig:lm2}, now including the first $1/N$ corrections with $N=6$.}
 \label{fig:lm2nlo}
\end{figure}
\begin{figure}[t]
 \begin{center}
  \includegraphics[width=.47\textwidth,angle=0]{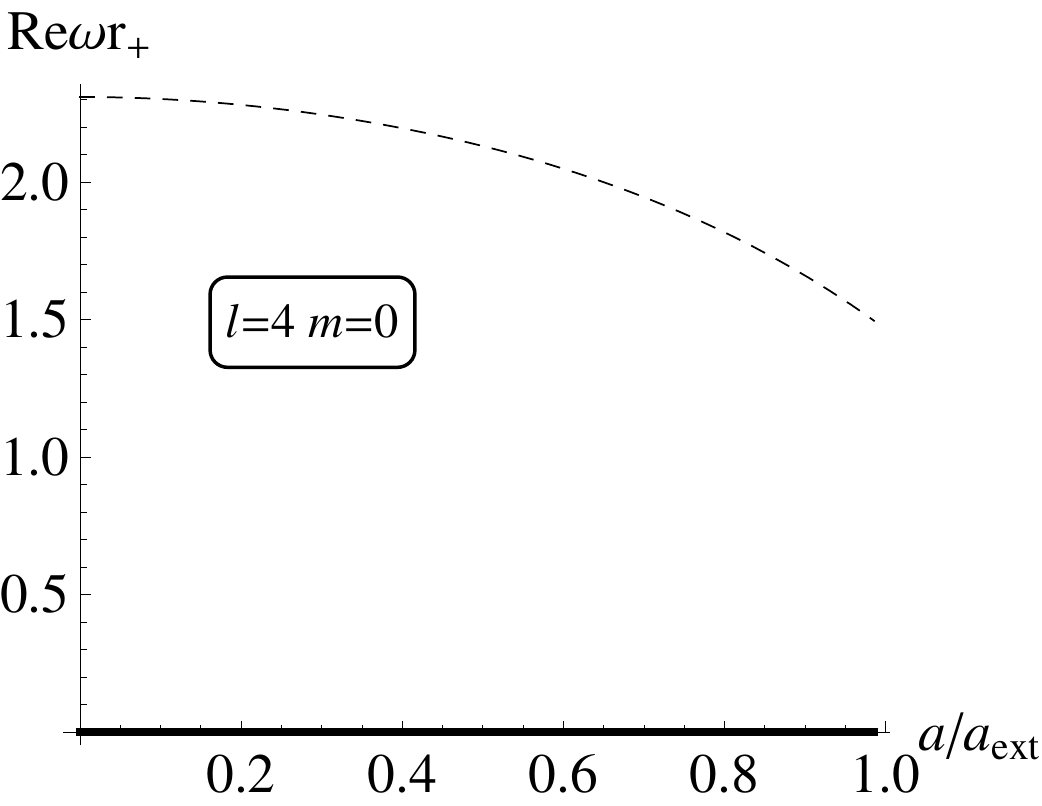}
  \hspace{5mm}
  \includegraphics[width=.47\textwidth,angle=0]{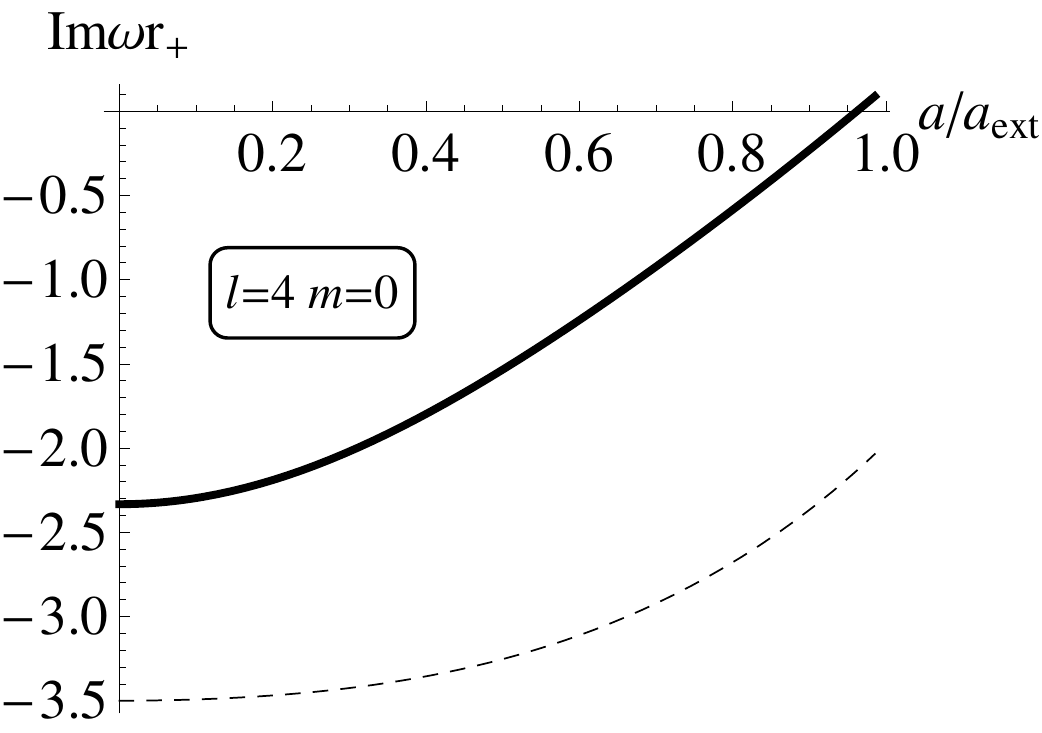}
   \end{center}
 \vspace{-5mm}
 \caption{\small Same as fig.~\ref{fig:l4m0}, now including the first $1/N$ corrections with $N=6$. We omit $\omega^{(-)}_{\ell,0}$ and $\omega^{(1)}_{\ell,0}$. Observe that the corrections eliminate the abrupt behavior of the leading-order result close to extremality.}
 \label{fig:l4m0nlo}
\end{figure}

Figs.~\ref{fig:lm2nlo} and \ref{fig:l4m0nlo} show the same quasinormal modes as in figs.~\ref{fig:lm2}, \ref{fig:l4m0} (except for $\omega^{(1)}_{\ell,0}$), now including the $1/N$ corrections with $N=6$. 

Since the coefficients of the corrections include factors of the type $\sim \ln\lp 1-(a/a_\text{ext})^2\rp$, we expect that our results for the critical rotation are more accurate for the bar-mode $\ell=m=2$, which has lower $a_c$, than for the axisymmetric mode with $\ell=4$.

\subsection{Comparison to numerical results}
\label{subsec:comp}

If we compare our result \eqref{acaext} to the numerical calculations of \cite{Hartnett:2013fba} for $N=2,\dots,6$, we find that we recover the critical rotations $a_c/a_\text{ext}$ with an accuracy that is typically better than $\approx 1/(2N)^2=1/(D-3)^2$. However, for modes with $m= 0$, for which the critical rotation is closer to extremality, the accuracy is rather $\approx 1/N^2$.
In table~\ref{table} we compare our calculations for the dominant unstable mode $\ell=m=2$ to the values of \cite{Hartnett:2013fba}.

\begin{table}[ht]
\begin{center}
\caption{\small Values for the critical rotation $a_c/a_\text{ext}$ for the dominant unstable mode $\ell=m=2$ in $D=2N+3=7,9,11,13,15$. The first row are the numerical values of \cite{Hartnett:2013fba} and the second row our analytical values from \eqref{acaext}. The third row gives the rotation at the threshold of the (thermodynamically-defined) ultraspinning regime: bar-mode instabilities appear \textit{before} this threshold is crossed. For reference, the leading order result is $a_c/a_\text{ext}=1/\sqrt{2}=0.7071$.}\label{table}
\medskip
\begin{tabular}{|c||c|c|c|c|c|}
 \hline
$N$&2&3&4&5&6\\
\hline
$a_c/a_\text{ext}  ~\cite{Hartnett:2013fba}$&0.8109&0.7463&0.7413&0.7369&0.7331\\
\hline
$a_c/a_\text{ext}~\eqref{acaext}$&0.7955&0.7660&0.7513&0.7425&0.7366\\
\hline
$a_\text{ultra}/a_\text{ext}$&0.8660&0.8165&0.7906&0.7746&0.7638\\
\hline
\end{tabular}
\end{center}
\end{table}
Fig.~\ref{fig:comp} shows that the quasinormal frequency as a function of $a$ is also very well reproduced --- even up to the extremal rotation $a_\text{ext}$.
%
%

\section{AdS rotating black hole} 
\label{sec:adsmp}

The previous analysis can be generalized straightforwardly to AdS rotating black holes with equals spins. Their metric takes the same form as \eqref{MPbh}
but now
%
\beq
G(r) =  1-\left(\frac{r_{0}}{r}\right)^{2N}\lp 1-\frac{a^2}{r^2}\rp +\frac{r^{2}}{L^{2}}H(r)
\,, 
\eeq
with $H(r)$ and $\Omega(r)$ unchanged.
The horizon is again at 
\beq
r_+=r_0+\mc{O}(N^{-1})\,,
\eeq 
so we will interchangeably employ either of the two radii.
The rotation parameter $a$ is bounded above by the extremal limit, which at large $N$ is
%
\begin{eqnarray}
\frac{a_{\text{ext}}}{r_+} = \frac{1}{\sqrt{1+r_+^2/L^{2}}}+\mc{O}(N^{-1})\,.
\end{eqnarray}
%
More properties of these black holes are given in appendix~\ref{app:hessian}.

\subsection{Large $D$ limit} 

We introduce the near-horizon coordinate
\begin{eqnarray}
\sR = \lp\frac{r}{r_0}\rp^{2N}\left( 1-\frac{a^{2}}{r^2}\lp 1+\frac{r^{2}}{L^{2}} \rp\right)^{-1}\,.
\end{eqnarray}
For the time being we keep the parameter $r_0$.
In terms of $\sR$ the leading order near-horizon geometry is
%
\begin{eqnarray}
ds^{2} &=&\left( 1+\frac{r_+^2}{L^{2}}-\frac{1}{\sR}\right)^{-1}\frac{r_+^2}{4N^{2}}\frac{d\sR^2}{\sR^2} 
-\left( 1+ \frac{r_+^2}{L^{2}} -\frac{\cosh^{2}{\alpha_{L}}}{\sR} \right)dt^{2} \notag \\
&&
-\frac{2r_+}{\sqrt{1+r_+^2/L^{2}}}\frac{\sinh{\alpha_{L}}\cosh{\alpha_{L}}}{\sR}dt(d\psi +A_{a}dx^{a})  \notag \\
&&
+r_+^2\left( 1+\frac{\sinh^{2}{\alpha_{L}}}{(1+r_+^2/L^{2})\sR} \right)(d\psi +A_{a}dx^{a})^{2}
+r_+^2\hat{g}_{ab}dx^{a}dx^{b},
\end{eqnarray}
%
where
%
\begin{eqnarray}
\tanh{\alpha_{L}} = \frac{a}{r_+}\,\sqrt{1+\frac{r_+^2}{L^2}}\,.
\end{eqnarray}
%
This metric is locally a frame transformation of the near-horizon geometry of the Schwarzschild-AdS black hole, of the form
%
\begin{gather}
dt \rightarrow dt \cosh{\alpha_{L}} -(d\psi+A_{a}dx^{a}) \frac{r_+\sinh{\alpha_{L}}}{\sqrt{1+r_+^2/L^{2}}}, \\
d\psi+A_{a}dx^{a} \rightarrow (d\psi+A_{a}dx^{a}) \cosh{\alpha_{L}} -\frac{dt}{r_+} \sqrt{1+\frac{r_+^2}{L^2}}\,\sinh{\alpha_{L}}.
\end{gather}
%
This implies that rotating AdS frequencies that only probe the leading order geometry can be derived from Schwarzschild-AdS frequencies in the form
\beq\label{AdStrans1}
\omega =\frac{m}{r_+}\,\sqrt{1+\frac{r_+^2}{L^2}}\, \tanh{\alpha_{L}} +\omega_\text{Sch-AdS}\, \text{sech}\,{\alpha_{L}}\,.
\eeq

Furthermore, the effects of the AdS cosmological constant on the metric amount to the transformations
\beq
(t,\psi,x^a)\to(t,\psi,x^a)\sqrt{1+\frac{r_+^2}{L^2}}\,,\qquad \sR\to \sR\lp 1+\frac{r_+^2}{L^2}\rp\,,\qquad r_+\to r_+\sqrt{1+\frac{r_+^2}{L^2}}
\eeq
and
\beq\label{AdSrescale}
a\to a\,\sqrt{1+\frac{r_+^2}{L^2}}\,.
\eeq
This implies that for those modes that are sensitive only to the leading order geometry, the frequencies of Schwarzschild or MP quasinormal modes are mapped to frequencies for AdS black holes by changing $a$ as above and
\beq \label{AdSrescale2}
\omega r_+\to\omega r_+\,,\qquad m\to m\,\sqrt{1+\frac{r_+^2}{L^2}}\,.
\eeq

Putting together the two observations, we get that quasinormal modes of Schwarzschild (without cosmological constant) yield frequencies of the rotating AdS black hole as
\beq\label{schtompads}
\omega r_+= m\frac{a}{r_+}\lp 1+\frac{r_+^2}{L^2}\rp  +\omega_\text{Sch}r_+\sqrt{1-\frac{a^2}{r_+^2}-\frac{a^2}{L^2}}\,.
\eeq
Below we will confirm this rule by explicit calculation of certain quasinormal frequencies. However, the same caveat applies: mode frequencies that depend on the geometry beyond leading order are not obtained in this straightforward manner. 

Henceforth we set $r_0=1$ so 
\beq
r_+=1+\mc{O}(N^{-1})\,.
\eeq

\subsection{Perturbations and quasinormal modes}

Using 
the frame transformations we can construct decoupled variables, in terms of which we have
%
\begin{eqnarray}
&&
f_{00}= 
\frac{L^{2}\sR(F_{00}+F_{11}) -2a(L^{2}(\sR-1)+\sR )F_{02} +a^{2}(L^{2}(\sR-1)+\sR)F_{22}  }
{L^{2}\sR -a^{2}(L^{2}(\sR-1)+\sR)},
\end{eqnarray}
%
%
\begin{eqnarray}
f_{01} = \frac{L^{2}\sR F_{01} -a(L^{2}(\sR-1)+\sR)F_{12}}{L\sqrt{\sR(L^{2}\sR-a^{2}(L^{2}(\sR-1)+\sR))}},
\end{eqnarray}
%
%
\begin{eqnarray}
f_{02} &=&\sqrt{\frac{L^{2}(\sR-1)+\sR}{\sR}} \notag \\
&&\times
\frac{(L^{2}\sR+a^{2}(L^{2}(\sR-1)+\sR))F_{02} -aL^{2}\sR(F_{00}+F_{11}+F_{22})}{L(L^{2}\sR-a^{2}(L^{2}(\sR-1)+\sR))},
\end{eqnarray}
%
%
\begin{eqnarray}
f_{12} = \frac{\sqrt{L^{2}(\sR-1)+\sR}(F_{12}-aF_{01})}{\sqrt{L^{2}\sR-a^{2}(L^{2}(\sR-1)+\sR)}},
\end{eqnarray}
%
%
\begin{eqnarray}
f_{22}=
\frac{L^{2}\sR F_{22} +a^{2}(L^{2}(\sR-1)+\sR)(F_{00}+F_{11})-2a(L^{2}(\sR-1)+\sR)F_{02} }{L^{2}\sR -a^{2}(L^{2}(\sR-1)+\sR)},
\end{eqnarray}
%
%
\begin{eqnarray}
f_{0}^{+} = \frac{L^{2}\sR F_{0} -a(L^{2}(\sR-1)+\sR)F_{2}}{L\sqrt{\sR(L^{2}\sR-a^{2}(L^{2}(\sR-1)+\sR))}},
\end{eqnarray}
%
%
\begin{eqnarray}
f_{2}^{+} = \frac{\sqrt{L^{2}(\sR-1)+\sR}(F_{2}-aF_{0})}{\sqrt{L^{2}\sR-a^{2}(L^{2}(\sR-1)+\sR)}},
\end{eqnarray}
%
and
%
\begin{eqnarray}
f_{11} =F_{00}-F_{11}\,,\qquad
f_{1}^+=F_{1}\,.
\end{eqnarray}
%
The decoupling variables are
%
\begin{eqnarray}
F_{AB}\,,\,~F_{A}\,,\,~H^{+-}.
\end{eqnarray}
%

The boundary conditions  at $\sR\gg 1$ follow from the Dirichlet boundary conditions in the asymptotic far zone, which implies that perturbations at large $\sR$ behave as $\sim \sR^{-1}$. This is the same as we had for the MP black hole. Everything else is entirely analogous to what we have done above, so we only quote the final results.

\paragraph{Leading order.}
The leading order solution satisfying the required boundary conditions contains integration constants $A_{0}$, $C_{0}$,
$D_{0}$, $G_{0}$ like in the case of the MP black hole.

\paragraph{Next-to-leading order.}
For perturbations with $A_{0}\neq 0$ we obtain a quasinormal frequency
%
\begin{eqnarray}
\omega^{(1)}_{(\ell,m)} &=& am\left(1+\frac1{L^{2}}\right) -i\ell\sqrt{1-a^{2}-\frac{a^2}{L^2}} \,.
\label{AdSQNM1}
\end{eqnarray}
%
and $G_{0}=0$. This result can also be obtained from the MP frequency eq.~(\ref{qnm1}) by rescaling the parameters in the form
(\ref{AdSrescale}--\ref{AdSrescale2}). It also follows at once from the Schwarzschild frequency \eqref{Schv} using \eqref{schtompads} and the vector transformation rule $\ell=\ell_{\text{sph}}-1$.

\paragraph{Next-to-next-to-leading order.} We first find a regular mode with frequency
\begin{eqnarray}\label{ome0}
\omega = am\left(1+\frac1{L^{2}}\right) -i\lp \ell+\frac2{L^2}\rp\sqrt{1-a^{2}-\frac{a^2}{L^2}}\,,
\end{eqnarray}
%
which at finite $L$ is different than \eqref{AdSQNM1}. However, this is again a gauge mode.

We obtain the equation for physical quasinormal frequencies as
%
\begin{eqnarray}
&&
\frac{1}{\omega L^{2}-a(1+L^{2})(m+2)+i(\ell -2)L\sqrt{L^{2}-a^{2}(1+L^{2})} } \notag \\
&&
\times \Bigl[
-i L^{9}\sqrt{L^2-a^2 \left(1+L^2\right)} \omega^{3} \notag \\
&&~~~~~
-\Bigl( (4-3 \ell )L^3+a^2 (3 \ell -4) \left(1+L^2\right) L  \notag \\
&&~~~~~~~~~~~~~~~~~~~~~~~~~~~~~~
-3 i a \left(L^2+1\right)\sqrt{L^2-a^2 \left(1+L^2\right)} m\Bigr) L^{7}\omega^{2} \notag \\
&&~~~~~
+\Bigl( 6(\ell -1) L \left(1+L^2\right)^2 m a^3 -i\left(1+L^2\right)\sqrt{L^2-a^2 \left(1+L^2\right)}  \notag \\
&&~~~~~~~
\times
\left(3 \ell (\ell -2)L^2+3\left(1+L^2\right) m^2+\ell -4\right) a^2 -6 (\ell -1) L^3 \left(1+L^2\right)m a\notag \\
&&~~~~~~~~~~~~~~
+i L^2 \left((\ell -1) (3 \ell -4) L^2+\ell \right) 
\sqrt{L^2-a^2\left(1+L^2\right)}\Bigr)L^{5} \omega \notag \\
&&~~~~
+\bigl( (2-\ell ) \ell  \left((\ell -1) L^2+1\right) L^5+a^2 \left(1+L^2\right) 
\bigl( (3 \ell -2)\left(1+L^2\right) m^2 \notag \\
&&~~~~
+\ell (\ell -2) \left( (2 \ell -1) L^2+2\right)\bigr) L^3 
 -i a \left(1+L^2\right) \left((\ell -1) (3 \ell -2) L^2+\ell +2\right) \notag \\
&&~~~~
\times
\sqrt{L^2-a^2 \left(1+L^2\right)} m L^2 +a^4 \left(1+L^2\right)^2
\bigl(\ell (2-\ell ) \left(\ell  L^2+1\right) \notag \\
&&~~~~~
-(3 \ell -2) \left(1+L^2\right)m^2\bigr) L +i a^3 \left(1+L^2\right)^2 \sqrt{L^2-a^2
   \left(1+L^2\right)} m \notag \\ 
&&~~~~~
\times\left(\ell (3 \ell -4 ) L^2+\left(1+L^2\right)m^2+\ell -2\right)\bigr) L^3
\Bigr] =0. \label{AdSqnmc}
\end{eqnarray} 
%

The properties of these perturbations of rotating AdS black holes can be obtained from \eqref{AdSqnmc} like we did for the MP black hole, with results that are entirely analogous, so we will be brief.

At the critical rotation
%
\begin{eqnarray}
a &=& a_c = \sqrt{\frac{\ell-1+1/L^{2}}{(1+1/L^{2})(\ell +1/L^{2})}} \notag \\
&=& a_{\text{ext}}\sqrt{\frac{\ell-1+1/L^{2}}{\ell +1/L^{2}}},
\end{eqnarray}
%
there is a mode with
\beq
\text{Im}\,\omega =0,\qquad \text{Re}\,\omega=am\lp 1+\frac1{L^2}\rp\,,
\eeq
which sits at the superradiance bound.
For $a>a_c$ the mode becomes unstable and superradiant. For $\ell=2$ the critical rotation is the same as the ultraspinning bound~\eqref{ultradef}.

When $m=0$ there is a mode with purely imaginary frequency, which is the one that, for $\ell\geq 4$, becomes unstable.
For $\ell=2$ and $m=0$ the three independent frequencies take the simple form
%
\beq
\omega =\begin{cases}0\,,\\ \pm\sqrt{\lp 1+\frac{2}{L^{2}}\rp\lp 1+a^{2}+\frac{a^2}{L^{2}}\rp}-i\sqrt{ 1-a^{2}+\frac{a^2}{L^{2}}}\,.
\end{cases}
\eeq
%
$\omega=0$ corresponds to a variation of the angular momentum, so there is no unstable mode with $(\ell,m)=(2,0)$. Axisymmetric instabilities begin at $a_c$ with $\ell=4$.

\begin{figure}[t]
 \begin{center}
  \includegraphics[width=.47\textwidth,angle=0]{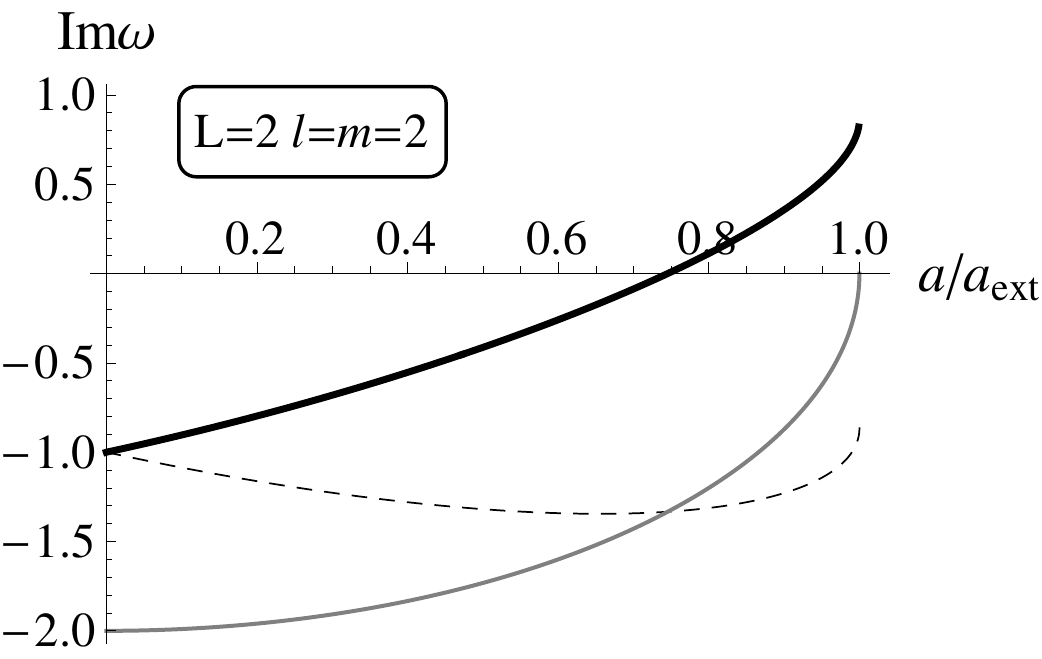}
  \hspace{5mm}
  \includegraphics[width=.47\textwidth,angle=0]{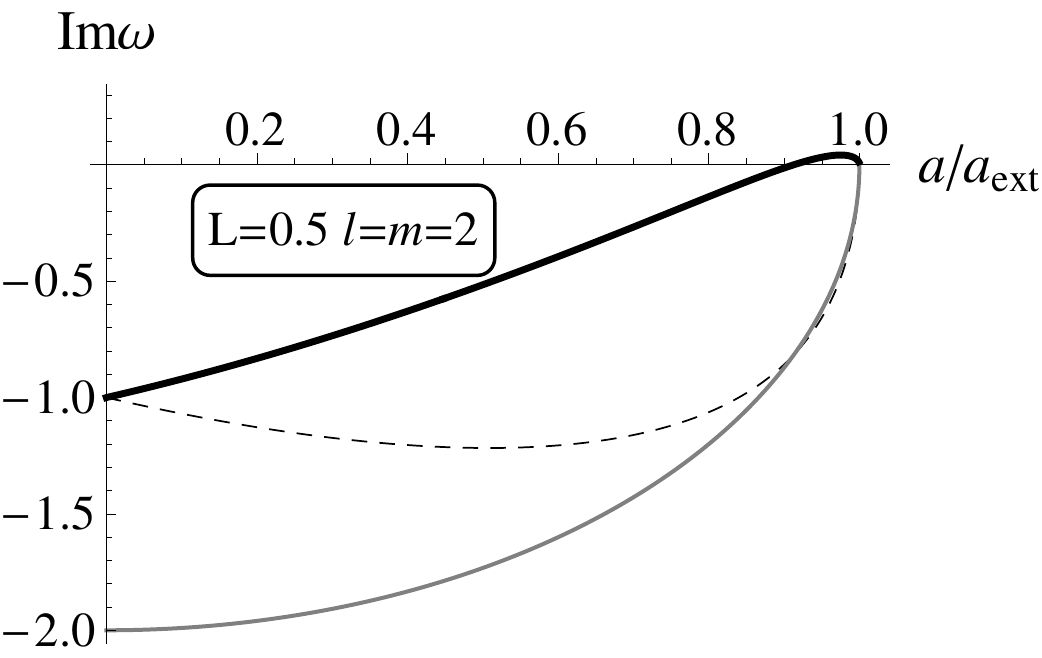}
   \end{center}
 \vspace{-5mm}
 \caption{\small Imaginary frequencies of quasinormal modes $(\ell,m)=(2,2)$ for (left) a relatively small AdS black hole ($L=2$, \ie\ $r_+/L=1/2$), and (right) a relatively large one ($L=0.5$, \ie\ $r_+/L=2$). For comparison, fig.~\ref{fig:lm2} is the limit $L\to\infty$. The thick solid line is the mode that becomes unstable at $a_c/a_\text{ext}=\sqrt{(L^2+1)/(2L^2+1)}=0.745~(\text{left})\; 0.913~(\text{right})$. The instability is clearly suppressed as $L$ decreases.}
 \label{fig:Imlm2L}
\end{figure}

The behavior of quasinormal modes at different values of $L$ are qualitatively very similar to what we found in the absence of a cosmological constant. In order to illustrate the effect of finite $L$ on the instability we plot in fig.~\ref{fig:Imlm2L} the imaginary part of the frequency of the dominant unstable mode, with $\ell=m=2$, for a relatively small black hole and for a larger one. Fig.~\ref{fig:lm2} corresponds to the limit $L/r_+\to\infty$ of very small black hole. As $L$ decreases, which enhances the effect of the cosmological constant on the black hole, the unstable behavior is suppressed. Apparently, when $L$ is large $\text{Im}\,\omega$ remains finite in the extremal limit while at small $L$ (large black holes) it vanishes in that limit. However, it is unclear how significant or reliable this is, as our methods break down in the extremal limit. 

Although the choices of axes and of $L$ are not the same, our fig.~\ref{fig:Imlm2L}\,(right) can be compared with fig.~18\,(right) of \cite{Cardoso:2013pza}. The qualitative agreement for the mode behavior is manifest. This occurs also for the real parts of the frequencies, as well as for other modes.

\subsection{Limit $a=0$ and black brane hydrodynamic limit}

Our discussion in sec.~\ref{subsec:static} of the limit $a=0$ extends in the same qualitative manner. The AdS Schwarzschild black hole also has quasinormal frequencies $\omega=\mc{O}(D^0)$ of scalar- and vector-type, which we have computed in the formalism of \cite{Kodama:2003jz} with the result that
%
\beq\label{AdSs}
\omega^{\text{scalar}}_{(\pm)} = \pm \sqrt{\ell_{\text{sph}}-1 +\frac{\ell_{\text{sph}}}{L^{2}}} -i(\ell_{\text{sph}}-1)
\eeq
%
and
%
\begin{eqnarray}
\omega^{\text{vector}} = -i(\ell_{\text{sph}}-1)\,. \label{AdSv}
\end{eqnarray}
%
The relation of these modes to those of the previous subsection in the limit $a=0$ is completely analogous to the description in sec.~\ref{subsec:static}. In particular,
applying the transformation rule \eqref{AdStrans1} to $\omega^{\text{vector}} $ with $\ell_\text{sph}=\ell+1$ we obtain \eqref{AdSQNM1}. 

Taking $L\to \infty$ in \eqref{AdSs}, \eqref{AdSv} reproduces the results \eqref{Schs}, \eqref{Schv} for Schwarzschild. On the other hand, in the limit $L\to 0$, \ie\ $r_+/L\to\infty$, which corresponds to very large AdS black holes limiting to AdS black branes, we obtain the hydrodynamic modes of black branes once we identify the momentum $q$ along the brane as
\beq
\frac{\ell_{\text{sph}}}{L^2}\to \frac{q^2}{D}\,.
\eeq
Then $\omega_{(\pm)}^{\text{scalar}}$ become sound modes and $\omega^{\text{vector}}$ a shear mode.


\section{High frequency modes: $\omega =\mc{O}(D)$}
\label{sec:odmodes}

Up to now we have focused on quasinormal modes with $\omega,\,\ell =\mc{O}(D^0)$, since these are the ones where the instabilities occur. However, there is a large class of other quasinormal modes with much higher frequency, $\omega,\,\ell =\mc{O}(D)$. These modes have been argued to be universally present for very large classes of static, spherically symmetric black holes, including the Schwarzschild solution, with complex frequencies \cite{Emparan:2014cia}
\beq\label{uniom}
\omega_{(\ell,k)}^\text{static}=\frac{D}2 +\ell -\lp \frac{e^{i\pi}}{2}\lp\frac{D}2 +\ell\rp\rp^{1/3}\mathsf{a}_k
\eeq
(with $r_0=1$), where $k=1,2,\dots$ is the `overtone' number and $-\mathsf{a}_k<0$ are zeroes of the Airy function Ai. These frequencies appear for gravitational scalars, vectors and tensors. In principle these are indexed by $\ell_\text{sph}$,  but to leading order at large $D$ only large values of $\ell_\text{sph}=\mc{O}(D)$ are relevant, so we can neglect the $\mc{O}(1)$ differences between $\ell$ and $\ell_\text{sph}$ when using $\mathbb{CP}^N$ harmonics.

The derivation of the spectrum \eqref{uniom} only relies on the leading order near-horizon geometry (in fact, it requires little information about it) and not on subleading structure. Therefore, it is possible to use the boost relation \eqref{schboost} to obtain the corresponding quasinormal spectrum for MP black holes, namely,\footnote{We have also gone in detail through the perturbation equations for the rotating black hole and checked explicitly that they are a boosted version of the ones for the Schwarzschild solution.}
\beq\label{unirot}
\omega_{(\ell,m,k)}=m a +\omega_{(\ell,k)}^\text{static}\sqrt{1-a^2}\,.
\eeq
The first term contributes at this order only when $m=\mc{O}(D)$.
A similar relation is expected to hold also for other classes of rotating black holes, not necessarily neutral ones. In the presence of the cosmological constant, the relation should apply to small AdS black holes, but the spectrum \eqref{uniom} is absent from large ones.

\section{Final remarks}
\label{sec:fin}

We have seen that the large $D$ analysis greatly simplifies the study of perturbations of a rotating black hole, making it analytically tractable. The main reasons for this are:
\begin{itemize}

\item The far-zone analysis is trivial.

\item The non-trivial dynamics in the near-horizon zone is simplified owing to the universality of the large $D$ limit of neutral black holes \cite{Emparan:2013xia}. This casts the perturbation problem of the rotating black hole into a boosted version of the static black hole problem. 

\item For some perturbations this universality allows to directly obtain the quasinormal spectrum via \eqref{schboost}. For other perturbations (including the unstable ones) further work is required but their equations decouple and are analytically solvable.

\end{itemize}

Since the universal near-horizon geometry of neutral black holes is also present for AdS black holes, the same reasoning applies to the study of their rotating quasinormal modes. A good example of the power of the relations between modes that follow from this universality is the fact that the rotating AdS black hole quasinormal frequency \eqref{AdSQNM1} can be obtained simply from knowledge of the Schwarzschild frequency \eqref{Schv}.

Our large $D$ analysis indicates the existence of an instability of these rotating black holes at sufficiently large $D$. However, this does not allow to deduce what is the lowest value of $D$ at which this conclusion applies, and in fact since the instabilities appear before the ultraspinning range \eqref{instwin}, this approach does not provide any indication of the absence of instability for $N=1$ \cite{Murata:2008yx}. This limitation is of course typical of perturbative methods. 

In a forthcoming article we will present the calculation of the quasinormal mode spectrum of Schwarzschild and Schwarzschild-AdS black holes in a more comprehensive way and to higher order in the expansion, using a different framework for the perturbations. 
It should also be interesting to extend the large $D$ study of stability to singly-spinning black holes.

\section*{Acknowledgments}

We are very grateful to \'Oscar Dias, Gavin Hartnett, and Jorge Santos for discussions and for supplying the numerical data used in fig.~\ref{fig:comp}.
Work supported by MEC FPA2010-20807-C02-02, AGAUR 2009-SGR-168 and CPAN CSD2007-00042 Consolider-Ingenio 2010. KT is supported by a grant for research abroad by JSPS.


\addcontentsline{toc}{section}{Appendices}
\appendix

\section{Hessian and ultraspinning surface for AdS rotating black holes}
\label{app:hessian}

Here we compute the Hessian 
\beq
H_{ij}=-\frac{\partial^2 S}{\partial J_i \partial J_j}
\eeq
of the rotating AdS black hole with equal angular momenta in $D=2N+3$ and identify its negative eigenvalues. This is sometimes called the reduced Hessian \cite{Dias:2010eu,Dias:2010maa} since it does not include variations in the mass of the solution. At low angular momenta this Hessian is positive definite. The codimension one surface that bounds this region in parameter space is called the ultraspinning surface. Beyond it, at sufficiently large angular momenta, the Hessian acquires negative eigenvalues, which correspond to thermodynamic negative modes.

Although we are interested in equal spin black holes, we want to study variations that do not leave all spins equal and therefore we need to work with the solution with unequal rotation parameters $a_i$, $i=1,\dots,N+1$. The relevant thermodynamic quantities have been computed in \cite{Gibbons:2004ai}. These solutions are written using a radial coordinate $\rho$ that in the equal-spin limit $a_i=a$ differs from the coordinate $r$ that we use; they are related by
\begin{eqnarray}
r^{2}=\frac{\rho^{2}+a^{2}}{\Xi}\,, \label{rdef}
\end{eqnarray}
where
\beq
\Xi=1-\frac{a^2}{L^2}\,,
\eeq
so the corresponding horizon radii are
\beq
r_+^{2}=\frac{\rho_+^{2}+a^{2}}{\Xi}\,.
\eeq
The physical magnitudes are expressed using a parameter $\rho_{0}$ which when all rotations are equal is related to our parameter $r_0$ by
%
\begin{eqnarray}
r_{0}^{2} =\frac{\rho_0^2}{\Xi^{1+2/N}}\,. \label{r0def}
\end{eqnarray}

For generic $a_i$ the horizon $\rho=\rho_{+}$ is located where
%
\begin{eqnarray}
(\rho_{+}^{2}+L^{2})\prod_{i=1}^{N+1}(\rho_{+}^{2}+a^{2}_{i})-L^{2}\rho_{+}^{2}\rho^{2N}_{0}=0.
\end{eqnarray}
%
The thermodynamic charges computed in \cite{Gibbons:2004ai} are the entropy
%
\begin{eqnarray}
S = \frac{\Omega_{D-2}}{4\pi \rho_{+}}\prod_{i=1}^{N+1}\frac{\rho_{+}^{2}+a^{2}_{i}}{\Xi_{i}},
\end{eqnarray}
mass
\begin{eqnarray}
M = \frac{\rho^{2N}_{0} \Omega_{D-2}}{8\pi \prod_{i=1}^{N+1}\Xi_{i}}\left( \sum_{i=1}^{N+1}\frac{1}{\Xi_{i}} -\frac{1}{2}\right),
\end{eqnarray}
%
and angular momenta
%
\begin{eqnarray}
J_{i} = \frac{a\rho_{0}^{2N}\Omega_{D-2}}{8\pi \Xi_{i}\prod_{j=1}^{N+1}\Xi_{j}}\,.
\end{eqnarray}
%
Here $\Xi_i=1-a_i^2/L^2$.
The surface gravity and angular velocity of the horizon are 
%
\begin{eqnarray}
\kappa = \rho_{+}\frac{\rho^{2}_{+}+L^{2}}{L^{2}}\sum_{i=1}^{N+1}\frac{1}{\rho_{+}^{2}+a^{2}_{i}} -\frac{1}{\rho_{+}}\,,\qquad
\Omega_{i} = \frac{a_{i}(\rho^{2}_{+}+L^{2})}{L^{2}(\rho^{2}_{+}+a^{2}_{i})}. 
\end{eqnarray}
%

When all angular momenta are equal the rotation parameter $a$ is bounded above by the extremal limit, defined by the condition that $\kappa=0$, \ie\ 
%
\begin{eqnarray}
(N+1)\rho^{2}_{+}(\rho^{2}_{+}+L^{2}) -L^{2}(\rho^{2}_{+}+a^{2}_\text{ext}) =0.
\end{eqnarray}
%
One can see that $a_\text{ext}<L$ for any $N\geq 1$.

We proceed to compute the Hessian. It takes the form
%
\begin{eqnarray}
H_{ij} = A\lp \delta_{ij}-\frac1{N+1}Q_{ij}\rp+\frac{B}{N+1} Q_{ij}\,, 
\end{eqnarray}
where $Q_{ij}=1$ for all $i,j$, and $A$ and $B$ are defined by
\beq
A=\frac{2a\pi}{\kappa J}\frac{(L^{2}+\rho^{2}_{+})(L^{2}-a^{2})(\rho^{2}_{+}-a^{2})}{L^{2}(L^{2}+a^{2})(a^{2}+\rho^{2}_{+})}\,,
\eeq
\beqa
B=&&\frac{2a\pi}{\kappa J}
\frac{L^{2}+\rho^{2}_{+}}{(L^{2}-a^{2})((2N+1)L^{2}-a^{2})\bigl( (N+1)\rho^{4}_{+}+NL^{2}\rho^{2}_{+}-a^{2}L^{2} \bigr)^{2}} \notag \\
&&~~~~~
\times
\Bigl[ 
a^{6}L^{2}+(2N+1)L^2\rho_{+}^2(NL^{2}+(N+1)\rho^{2}_{+})^2 +a^{4}\bigl( (5N+3)L^{4} \notag \\
&&~~~~~~~~~~~~
+(10N+9)L^{2}\rho^{2}_{+}+3(N+1)\rho^{4}_{+} \bigr) +a^{2}\bigl( N(2N+1)L^{6} \notag \\
&&~~~~~~~~~~~~
+(11N^{2}+12N+3)L^{4}\rho^{2}_{+}+3(N+1)(4N+3)L^{2}\rho^{4}_{+} \notag \\
&&~~~~~~~~~~~~
+3(N+1)^{2}\rho^{6}_{+} \bigr)
\Bigr]. \label{qdef}
\end{eqnarray}
%

For the equal-spin black hole there are two non-equivalent choices for the eigenvectors $V=(V_i)$ \cite{Dias:2010maa}.\footnote{
In the limit $L\rightarrow \infty$ we reproduce the eigenvalues of the MP black hole in \cite{Dias:2010maa}. 
Note however that eqs.(2.15) and (2.16) of that article appear to contain misprints.
} 
The first one is $V_{i}=v$ for all $i$, which describes deformations of the solution that preserve the equality of all angular momenta. The eigenvalue is $B$, which remains finite and positive 
in the range $0\leq a< a_\text{ext}$.

The other choice is such that $\sum_{i}V_{i}=0$, and represents an anisotropic deformation in an azimuthal direction. In this case the eigenvalue is $A$, which changes sign at the ultraspinning surface defined by
%
\begin{eqnarray}
a_{\text{ultra}} =\rho_{+}\,, 
\end{eqnarray}
%
\ie\
\beq
a_{\text{ultra}}=\frac{r_+}{\sqrt{2+\frac{r_+^2}{L^2}}}\,.\label{ultradef}
\eeq
This result is exact in $N$.

When $N$ is large, the horizon of the equal-spin solution is at
%
\begin{eqnarray}
\rho_{+} = \sqrt{\rho^{2}_{0}-a^{2}}+\mc{O}(N^{-1}), 
\end{eqnarray}
%
\ie\
\beq
r_{+}=r_{0}+\mc{O}(N^{-1})\,, 
\eeq
and the extremality bound is
%
\begin{eqnarray}
a_{\text{ext}}=\rho_{0}+\mc{O}(N^{-1}),
\end{eqnarray}
%
\ie\
%
\begin{eqnarray}
a_{\text{ext}} = \frac{r_{0}}{\sqrt{1+\frac{r^{2}_{0}}{L^2}}}+\mc{O}(N^{-1}).
\end{eqnarray}
%
To leading order this is the same as \eqref{adsext}. Note that $a_{\text{ext}}<L$ (even at finite $N$). When $r_+\gg L$ the extremal rotation $a_\text{ext}$ approaches $L$. Near this extremal limit we have $M\to a J\propto r_0^{2N}$, so $r_0$ is a more convenient parameter than $\rho_0$.

The extended Hessian that includes derivatives with respect to the mass has no negative eigenvalues at sufficiently high $r_+/L$ and low $a/r_+$ \cite{Dias:2010gk}. This is a regime where the rotating black hole is completely thermodynamically stable, but our analysis of the reduced Hessian cannot capture it.

\section{Harmonics of $S^{2N+1}$ and $\mathbb{CP}^N$}
\label{app:sncpn}

In this article we are writing the metric of $S^{2N+1}$ as a $U(1)$ fibration of $\mathbb{CP}^N$,
\beq
ds^2_{S^{2N+1}}=\lp d\psi +A_a dx^a\rp^2 +\hat g_{ab}dx^a dx^b\,.
\eeq
Harmonics of $S^{2N+1}$ can then be written in terms of harmonics of $\mathbb{CP}^N$. We will be interested in scalar and vector harmonics only, as there are no tensor perturbations with frequencies $\omega=\mc{O}(D^0)$.

Scalar harmonics $\mathbb{S}$ of $S^{2N+1}$ satisfy
\beq
\left[ \nabla^2+\ell_\text{sph}(\ell_\text{sph}+2N))\right] \mathbb{S}=0\,,
\eeq
with $\nabla^2$ the Laplacian on the sphere and $\ell_\text{sph}=0,1,2,\dots$.
Vector harmonics $\mathbb{V}_\mu$ satisfy
\beq\label{vharm}
\left[ \nabla^2+\ell_\text{sph}(\ell_\text{sph}+2N)-1\right]  \mathbb{V}_\mu=0\,,\qquad \nabla^\mu\mathbb{V}_\mu=0 \,.
\eeq
Out of a scalar $\mathbb{S}$ we can construct a scalar-derived vector $\nabla_\mu \mathbb{S}$, which satisfies
\beq\label{svharm}
\left[ \nabla^2+\ell_\text{sph}(\ell_\text{sph}+2N)-2N\right]\nabla_\mu \mathbb{S}=0\,.
\eeq

Let us now consider $\mathbb{CP}^N$ harmonics of scalar and scalar-derived type, as presented in sec.~\ref{subsec:cpnharm}. From \eqref{cpnscal} it is easy to see that if $\mathbb{Y}$ is a $\mathbb{CP}^N$ scalar then
\beq
\nabla^2(e^{i m\psi}\mathbb{Y})=-\ell(\ell+2N)e^{i m\psi}\mathbb{Y}
\eeq
and therefore $e^{i m\psi}\mathbb{Y}$ is a scalar of $S^{2N+1}$ with $\ell_\text{sph}=\ell$.

Out of the $\mathbb{CP}^N$ scalar $\mathbb{Y}$ and scalar-derived vectors $\mathbb{Y}^\pm_a$ we can construct vectors $\mathbb{X}_\mu$ of $S^{2N+1}$ as linear combinations
\beq
\mathbb{X}_\mu=e^{i m\psi}\mathbb{Y}\, x_2\, e^{(2)}_\mu+e^{im\psi}\lp x_+\mathbb{Y}^+_a +x_-\mathbb{Y}^-_a \rp \hat e^{a}_{(i)} e^{(i)}_\mu
\eeq
with coefficients $x_2$ and $x_\pm$. The vielbein basis is
\beq
e^{(2)}=d\psi+A\,,\qquad e^{(i)}=\hat e^{(i)}
\eeq
with $\hat e^{(i)}$ a $\mathbb{CP}^N$ vielbein, and its inverse
\beq
e_{(2)}=\partial_\psi\,, \qquad e_{(i)}=\hat e^a_{(i)}(\partial_a -A_a\partial_\psi)\,.
\eeq
We choose three specific $S^{2N+1}$ vectors $\mathbb{X}_{\mu}^{(sv,v_+,v_-)}$ by taking
\begin{align}
&x_2^{(sv)}=im\,, & x_\pm^{(sv)}&=-\sqrt{\lambda}\,,\notag\\
&x_2^{(v_+)}=i\,, & x_\pm^{(v_+)}&=\mp\frac{\sqrt{\lambda}}{\ell+2N\pm m}\,,\\
&x_2^{(v_-)}=i(\ell^2-m^2)\,, & x_\pm^{(v_-)}&=\mp\sqrt{\lambda}(\ell\pm m)\,,
\notag
\end{align}
with $\lambda$ defined in \eqref{deflambda}. 
Then one finds that $\mathbb{X}_\mu^{(sv)}$ is a scalar-derived vector of $S^{2N+1}$  that satisfies \eqref{svharm} with $\ell_\text{sph}=\ell$, 
and $\mathbb{X}_\mu^{(v_\pm)}$ are vector harmonics that satisfy \eqref{vharm} with 
$\ell_\text{sph}=\ell\pm 1$, respectively. Note that $\mathbb{X}_\mu^{(v_-)}$ does not exist for $\ell=|m|$.

\section{Decoupling variables}
\label{app:decoupl}

The vielbein for the near-horizon MP solution is
%
\begin{eqnarray}
e^{(0)} = \sqrt{\frac{\sR-1}{\sR+\sinh^{2}{\alpha}}}dt\,,\qquad e^{(1)}=\frac{dr}{2N\sqrt{\sR(\sR-1)}},
\end{eqnarray}
%
%
\begin{eqnarray}
e^{(2)}=\left(1 +\frac{\sinh^{2}{\alpha}}{\sR}\right)^{1/2}\Bigl[ 
d\psi + A_{a}dx^{a}
-\left(1 +\frac{\sinh^{2}{\alpha}}{\sR}\right)^{-1}\sinh{\alpha}\cosh{\alpha}\sR^{-1}dt
\Bigr],
\end{eqnarray}
%
and $e^{(i)}$ the same as in the Schwarzschild black hole. The boost transformation (\ref{transL1}) and (\ref{transL2}) acts as
%
\begin{gather}
e^{(0)}_{\text{(Sch)}} 
\rightarrow e^{(0)}\sqrt{\frac{R}{R+\sinh^{2}{\alpha}}}\cosh{\alpha} -e^{(2)}\sqrt{\frac{R-1}{R+\sinh^{2}{\alpha}}}\sinh{\alpha}, \\
e^{(2)}_{\text{(Sch)}} 
\rightarrow e^{(2)}\sqrt{\frac{R}{R+\sinh^{2}{\alpha}}}\cosh{\alpha} -e^{(0)}\sqrt{\frac{R-1}{R+\sinh^{2}{\alpha}}}\sinh{\alpha}.
\end{gather}

The perturbation variables for the MP black hole are defined 
as in \eqref{hdef} using this vielbein.
Then the MP perturbation variables can be written in terms of the Schwarzschild ones as
%
\begin{eqnarray}
f_{00}=f_{00}^{(\text{Sch})} \frac{\sR}{\sR-a^{2}(\sR-1)}
-f_{02}^{(\text{Sch})} \frac{2a\sqrt{\sR(\sR-1)}}{\sR-a^{2}(\sR-1)} 
+f_{22}^{(\text{Sch})} \frac{a^{2}(\sR-1)}{\sR-a^{2}(\sR-1)}, 
\end{eqnarray}
%
%
\begin{eqnarray}
f_{01}=
f^{(\text{Sch})}_{01}\sqrt{\frac{\sR}{\sR-a^{2}(\sR-1)}}-af^{(\text{Sch})}_{12}\sqrt{\frac{\sR-1}{\sR-a^{2}(\sR-1)}},
\end{eqnarray}
%
%
\begin{eqnarray}
f_{02}=-f^{(\text{Sch})}_{00} \frac{a\sqrt{\sR(\sR-1)}}{\sR-a^{2}(\sR-1)}
+f^{(\text{Sch})}_{02} \frac{\sR+a^{2}(\sR-1)}{\sR-a^{2}(\sR-1)} 
+f^{(\text{Sch})}_{22} \frac{a\sqrt{\sR(\sR-1)}}{\sR-a^{2}(\sR-1)},
\end{eqnarray}
%
%
\begin{eqnarray}
f_{12}=
f^{(\text{Sch})}_{12}\sqrt{\frac{\sR}{\sR-a^{2}(\sR-1)}}-af^{(\text{Sch})}_{01}\sqrt{\frac{\sR-1}{\sR-a^{2}(\sR-1)}},
\end{eqnarray}
%
%
\begin{eqnarray}
f_{22}=f_{00}^{(\text{Sch})} \frac{a^{2}(\sR-1)}{\sR-a^{2}(\sR-1)}
-f_{02}^{(\text{Sch})} \frac{2a\sqrt{\sR(\sR-1)}}{\sR-a^{2}(\sR-1)} 
+f_{22}^{(\text{Sch})} \frac{\sR}{\sR-a^{2}(\sR-1)}, 
\end{eqnarray}
%
%
\begin{eqnarray}
f_{0}^+= 
f^{(\text{Sch})+}_{0}\sqrt{\frac{\sR}{\sR-a^{2}(\sR-1)}}-af^{(\text{Sch})+}_{2}\sqrt{\frac{\sR-1}{\sR-a^{2}(\sR-1)}}, 
\end{eqnarray}
%
%
\begin{eqnarray}
f_{2}^+ = 
f^{(\text{Sch})+}_{2}\sqrt{\frac{\sR}{\sR-a^{2}(\sR-1)}}-af^{(\text{Sch})+}_{0}\sqrt{\frac{\sR-1}{\sR-a^{2}(\sR-1)}},
\end{eqnarray}
%
The variables $f_{11}$, $f^{+}_{1}$ and $H^{+-}$ do not change form. 

The perturbations of Schwarzschild are decoupled in terms of the variables in \eqref{SchD}. This defines a set of decoupling variables for the MP solution too. We find convenient to define, as our decoupled variables, 
\begin{alignat}{3}
&F_{00} = \frac12\lp f^{\text{(Sch)}}_{00}+f^{\text{(Sch)}}_{11}\rp\,, &&\quad F_{01}=f^{\text{(Sch)}}_{01}\,, &&\quad F_{02}=\sqrt{\frac{\sR}{\sR-1}}f^{\text{(Sch)}}_{02} \,,\notag \\
&F_{11} = \frac12\lp f^{\text{(Sch)}}_{00}-f^{\text{(Sch)}}_{11}\rp\,, &&\quad F_{12}=\sqrt{\frac{\sR}{\sR-1}}f^{\text{(Sch)}}_{12}\,, &&\quad F_{22}=f^{\text{(Sch)}}_{22}, \\
&F_{0} = f^{\text{(Sch)}+}_{0}\,, &&\quad F_{1} = f^{\text{(Sch)}+}_{1}\,, &&\quad F_{2} = \sqrt{\frac{\sR}{\sR-1}} f^{\text{(Sch)}+}_{2} , \notag \\
&H^{+-}=H^{\text{(Sch)}+-}.&&&\notag
\label{FAB}
\end{alignat}
The expressions for the original variables $f_{AB}$, $f_A$ in terms of the decoupled ones are given in eqs.~(\ref{f00}--\ref{f11}).

\section{Stability of $\ell=0$ modes}
\label{app:ell0}

For modes with $\ell=m=0$ there are no scalar-derived vector nor tensor 
perturbations. Thus we consider only $f_{AB}$ and $H_{L}$. 
There is only one physical degree of freedom 
in this perturbation. We will derive the master equation for this mode by fixing the gauge explicitly, setting
%
\begin{eqnarray}
f_{00}=0\,,\qquad f_{02}=0\,,\qquad f_{22}=0.
\end{eqnarray}
%
In order to go to this gauge we transform with gauge function $\xi_{A}\sim f_{AB}/\omega$. Therefore these gauge conditions cannot be applied to the static perturbation $\omega=0$. However, the latter perturbation is simply a variation in the mass of the MP black hole, so we can ignore it. 

These gauge conditions yield secondary constraint equations, which correspond to the Einstein equations $G_{01}=0, G_{11}=0$ and
$G_{12}=0$. By solving them and using the equation for the trace part of $G_{ab}=0$, we obtain a master equation for $\ell=0$ perturbations,\footnote{Recall that we set $r_0=1$.}
%
\begin{eqnarray}
\left( \frac{d^{2}}{dr^{2}_{*}}+\omega^{2}-V(r) \right) \Psi_{0}=0,
\end{eqnarray}
%
where $r_{*}$ is the tortoise coordinate defined by $dr_{*}=\frac{\sqrt{H}}{G}dr$ and 
%
\begin{eqnarray}
\Psi_{0} = \frac{r^{N/2}(r^{2N+2}+a^{2})^{1/4}(r^{2N+2}-Na^{2})}{(2N+1)r^{2N+2}+Na^{2}}
H_{L}.
\end{eqnarray}
%
The potential $V(r)$ is given by 
%
\begin{eqnarray}
&&V(r)=\frac{G(r)}{4(r^{2N+2}+a^{2})^{3}((2N+1)r^{2N+2}+Na^{2})^{2}} \notag \\
&&~~~~~~~~~
\times\Bigl[ A_{0} +A_{1}r^{2}+A_{2}r^{2N+4}+A_{3}r^{4N+6}+A_{4}r^{6N+8}+A_{5}r^{8N+10} \Bigr],
\end{eqnarray}
%
where
%
\begin{eqnarray}
&&A_{0}=-a^{10}N^{3}(N+4), \\
&&A_{1}=a^{8}N^{2}(N^{2}-3(8+24N+11N^{2})r^{2N}), \\
&&A_{2}=-2a^{6}(-2N(8+31N+39N^{2}+17N^{3}) \notag \\
&&~~~~~~~~~~~~~~~~~~~
+(-16-78N-97N^{2}+4N^{3}+24N^{4})r^{2N}), \\
&&A_{3}=2a^{4}(2N+1)(N(28+59N+34N^{2}) \notag \\
&&~~~~~~~~~~~~~~~~~~~~~~~~~~
+(40+116N+65N^{2}+4N^{3})r^{2N}), \\
&&A_{4}=a^{2}(2N+1)(4(-1+5N+15N^{2}+10N^{3}) \notag \\
&&~~~~~~~~~~~~~~~~~~~~~~~~~~~
+3(21+62N+44N^{2}+8N^{3})r^{2N}), \\
&&A_{5}=(2N+1)^{2}((2N+1)^{2}+(15+16N+4N^{2})r^{2N}).
\end{eqnarray}
%
$V(r)$ reduces to the one in \cite{Murata:2008yx} for $N=1$.
The positivity of this potential at $r\geq r_{+}$, which guarantees the stability against $\ell=0$ perturbations, can be shown in the same way as in \cite{Murata:2008yx}. Thus the large $D$ expansion is not needed to prove the stability of these modes.

\section{`Parameter' modes and residual gauge modes}
\label{app:gauge}

Among the perturbations of the black hole there are some that are physical but trivial --- they only vary the parameters in the MP solution --- and others that are unphysical modes that can be eliminated by a residual gauge transformation. We discuss them briefly here. The arguments are valid also for the AdS rotating black hole.  

\subsection{Variation of parameters of MP black hole}

The MP black hole is parametrized by its mass and angular momentum. A variation of these parameters is a stationary perturbation of the solution. Ref.~\cite{Dias:2010eu} finds, from an analysis at $r=\infty$, that these are $(\ell,m)=(0,0)$ or $(\ell,m)=(2,0)$ perturbations. The former is a mass variation, and the latter a variation of angular momenta. We have indeed found in sec.~\ref{sec:axisym} a zero mode for $(\ell,m)=(2,0)$. It is possible to check explicitly that this effects a variation in angular momenta by comparing to the result of making a small variation in the angular momenta of the exact MP solution. 
In the analysis in app.~\ref{app:ell0} we omitted the zero mode with $(\ell,m)=(0,0)$, so those results do not have any ambiguity from variation of parameters. 

Perturbations with $\ell>2$ do not contribute to variations of the parameters \cite{Dias:2010eu}. 
Thus the zero modes at $a=a_c$ with $\ell>2$ that we found in sec.~\ref{sec:axisym} are genuine 
signals for the appearance of new branches of black hole solutions. 

\subsection{Residual gauge transformations}

The TT gauge leaves some residual gauge freedom. Here we discuss its consequences in $	\ell>0$ modes. 

We write the generator of gauge transformations, $\xi_{\mu}$, in the form
%
\begin{eqnarray}
\xi_{\mu}dx^{\mu} = \bigl[ \xi_{A}e^{(A)} + r( \xi_{+}\mathbb{Y}^{+}_{a}+\xi_{-}\mathbb{Y}^{-}_{a} ) dx^{a} \bigr] e^{-i\omega t}e^{im\psi}. 
\end{eqnarray}
%
The TT gauge condition and transformation rules for each perturbation variable are given in \cite{Dias:2010eu}. 
Solving them we obtain the following residual gauge generators\footnote{As done for the perturbation variables, we can construct decoupling variables for the equation of $\xi$ and obtain the most general generator 
of residual gauge transformations.  
} 
%
\begin{eqnarray}
\xi_{0}= \frac{r_{0}}{N}\left( \frac{\xi^{(0)}_{r}}{\sqrt{(\sR-1)(\sR-a^{2}(\sR-1))}} +\mc{O}(N^{-1}) \right),
\end{eqnarray}
%
%
\begin{eqnarray}
\xi_{1} = \frac{r_{0}}{N}\left( \frac{\xi^{(0)}_{r}}{\sqrt{\sR(\sR-1)}}+\mc{O}(N^{-1}) \right),
\end{eqnarray}
%
%
\begin{eqnarray}
\xi_{2}=\frac{r_{0}}{N}\left( -\frac{a\xi^{(0)}_{r}}{\sqrt{\sR(\sR-a^{2}(\sR-1))}}+\mc{O}(N^{-1}) \right),
\end{eqnarray}
%
and
%
\begin{eqnarray}
\xi_{\pm} = \mc{O}(N^{-3/2}), 
\end{eqnarray}
%
for all $\ell>1$  perturbations. Here we solved the equation for $\xi$ at leading order, and imposed that the 
transformation preserves the boundary conditions for the perturbation. $\xi^{(0)}_{r}$ is an integration constant. 
Thus, for the $\ell>1$ modes we have only one residual gauge parameter. 
For $\ell=m=1$ we have an additional residual gauge parameter, $\xi^{(0)}_{-}$. In this case $\xi$ becomes
%
\begin{eqnarray}
\xi_{0}= \frac{r_{0}}{N}\left( \frac{\xi^{(0)}_{r}+\xi^{(0)}_{-}}{\sqrt{(\sR-1)(\sR-a^{2}(\sR-1))}}
+\mc{O}(N^{-1}) \right),
\end{eqnarray}
%
%
\begin{eqnarray}
\xi_{1} = \frac{r_{0}}{N}\left( \frac{\xi^{(0)}_{r}+\xi^{(0)}_{-}(2\sR-1)}{\sqrt{\sR(\sR-1)}}+\mc{O}(N^{-1}) \right),
\end{eqnarray}
%
%
\begin{eqnarray}
\xi_{2}=\frac{r_{0}}{N}\left( -\frac{a(\xi^{(0)}_{r}+\xi^{(0)}_{-})}{\sqrt{\sR(\sR-a^{2}(\sR-1))}}
+\mc{O}(N^{-1}) \right),
\end{eqnarray}
%
and
%
\begin{eqnarray}
\xi_{-} = \frac{r_{0}}{\sqrt{N}}\left(- 2\sqrt{2}\xi^{(0)}_{-} +\mc{O}(N^{-1}) \right). 
\end{eqnarray}
%
For $\ell=m$ modes, $\mathbb{Y}^{+}$ and $\mathbb{Y}^{++}$ vanish, so we do not have $\xi_{+}$. 
The appearance of $\xi^{(0)}_{-}$ can be understood easily. The boundary condition for $H^{\pm\pm}$ restricts the integration constants
of $\xi_{\pm}$
\footnote{The integration constants of $\xi_{0}$ and $\xi_{2}$ are determined by the boundary conditions of $f^{\pm}_{0}$ and $f^{\pm}_{2}$. 
}. For $\ell>1$ modes, there is no degree of freedom that generates a regular solution for $H^{\pm\pm}$ to leading order of $\xi$. However, for $\ell=m=1$ modes, 
we have $\xi_{-}$, but we do not have $H^{--}$ because $\mathbb{Y}^{--}$ vanishes for $\ell=m=1$ even though $\mathbb{Y}^{-}$ does not vanish. 
Then there is no constraint on the integration constant of $\xi_{-}$, which is $\xi^{(0)}_{-}$. Hence the perturbation with $\ell=m=1$ has an additional 
residual gauge mode. This is exactly the same conclusion as in \cite{Kodama:2003jz} for the Schwarzschild black hole.

To see the effects of this residual gauge freedom on the perturbed solution, it is enough to consider only the leading order solution. 
The residual generator $\xi$ transforms the integration constants in the leading order solution (\ref{LOsol}) as
%
\begin{eqnarray}\label{CDgauge}
C_{0} \rightarrow C_{0} + 4\xi^{(0)}_{r}\,,\qquad D_{0} \rightarrow D_{0} -4\xi^{(0)}_{r},
\end{eqnarray}
%
for $\ell>1$, and
%
\begin{eqnarray}
C_{0} \rightarrow C_{0} +4\xi^{(0)}_{r}-4\xi^{(0)}_{-}\,\qquad D_{0} \rightarrow D_{0} -4\xi^{(0)}_{r}-4\xi^{(0)}_{-},
\end{eqnarray}
%
for $\ell=m=1$. $A_{0}$ and $G_{0}$ are not transformed: they are gauge independent parameters in the TT gauge. 
Thus the choice between $A_{0}=0$ or $A_{0}\neq 0$ is a physical one. 

In sec.~\ref{sec:qnman} at NNLO with $A_{0}= 0$ we found a frequency $\omega =am -i \ell\sqrt{1-a^2}$. Substituting in eq.~\eqref{CDeq} we see that this mode has $C_0+D_0=0$. Then it can be eliminated by a gauge transformation of the type \eqref{CDgauge}. 

For $\ell=m=1$ we can always set $C_{0}=0$ and $D_{0}=0$ by choosing $\xi^{(0)}_{r}$ and $\xi^{(0)}_{-}$. Thus 
the $\ell=m=1$ zero mode perturbation with $A_{0}=0$ that we found in sec.~\ref{sec:naxi} is purely gauge.  This can be understood from the perturbations of the Schwarzschild black hole. As mentioned 
above, the perturbations with $A_{0}=0$ originate at $a=0$ in scalar-type (in $S^{D-2}$) perturbations of the Schwarzschild black hole 
with $\ell_{\text{sph}}=\ell$. The scalar-type perturbation with $\ell_{\text{sph}}=1$ is purely gauge \cite{Kodama:2003jz}. 
Thus, our result shows that this holds also for $a>0$. 

Also, as we mentioned in the main text, we cannot set $C_{0}=0$ and $D_{0}=0$ for the
$\ell=m=1$ mode with $A_{0}\neq 0$, since at NNLO this condition does not yield a regular solution. 

It might seem that we can set $C_{0}=0$ or $D_{0}=0$ for $\ell>1$ perturbations by a choice of $\xi^{(0)}_{r}$. 
This is correct for the perturbations with $A_{0}\neq 0$, but presumably not for $A_{0}=0$. Here we have considered the perturbative solutions and  $\xi$ only
to leading order. At next-to-leading order, the perturbation solution has additional conditions on the integration constants from 
the leading order solutions such as eq.~(\ref{CDeq}). If we set $D_{0}=0$ with $A_{0}=0$, we should have either $C_{0}=0$ or some specific $\omega$. 
However, if we use this specific $\omega$, we do not obtain a regular solution at next-to-next-to-leading order. 
Therefore we cannot freely set $C_{0}=0$ or $D_{0}=0$ for $\ell>1$ perturbations with $A_{0}=0$. This may imply
that the NLO or NNLO equation for $\xi$ also gives a non-trivial condition on $\xi^{(0)}_{r}$, and it prohibits using $\xi^{(0)}_{r}$ for 
setting $C_{0}=0$ or $D_{0}=0$.


\end{document}